\documentclass[12pt,dvips]{article}
\usepackage{rotating}
\usepackage{amsmath}
\usepackage{amsthm}
\usepackage{amsfonts}
\usepackage[dvips]{epsfig}
\usepackage{graphicx}
\usepackage{amssymb}
\usepackage{cancel}
\usepackage{pstricks} 
\usepackage{lscape}
\usepackage{color}
\usepackage{cite}
\newcommand{\beq}{\begin{equation}}
\newcommand{\eeq}{\end{equation}}
\newcommand{\ga}{\lower.7ex\hbox{$\;\stackrel{\textstyle>}{\sim}\;$}}
\newcommand{\la}{\lower.7ex\hbox{$\;\stackrel{\textstyle<}{\sim}\;$}}
\def\K{K{\"a}hler}

\begin{document}

\def\thefootnote{\fnsymbol{footnote}}

\begin{flushright}
{\tt KCL-PH-TH/2014-18}, {\tt LCTS/2014-18}, {\tt CERN-PH-TH/2014-076}  \\
{\tt ACT-5-14, UMN-TH-3335/14, FTPI-MINN-14/12} \\
\end{flushright}

\begin{center}
{\bf {\Large A No-Scale Inflationary Model to Fit Them All }}
\vspace {0.1in}
\end{center}

\vspace{0.05in}

\begin{center}{\large
{\bf John~Ellis}$^{a}$,
{\bf Marcos~A.~G.~Garc\'ia}$^{b}$,
{\bf Dimitri~V.~Nanopoulos}$^{c}$ \\ and \\
\vspace{0.1in}
{\bf Keith~A.~Olive}$^{b}$
}
\end{center}

\begin{center}
{\em $^a$Theoretical Particle Physics and Cosmology Group, Department of
  Physics, King's~College~London, London WC2R 2LS, United Kingdom;\\
Theory Division, CERN, CH-1211 Geneva 23,
  Switzerland}\\[0.2cm]
  {\em $^b$William I. Fine Theoretical Physics Institute, School of Physics and Astronomy,\\
University of Minnesota, Minneapolis, MN 55455, USA}\\[0.2cm]
{\em $^c$George P. and Cynthia W. Mitchell Institute for Fundamental Physics and Astronomy,
Texas A\&M University, College Station, TX 77843, USA;\\
Astroparticle Physics Group, Houston Advanced Research Center (HARC), Mitchell Campus, Woodlands, TX 77381, USA;\\
Academy of Athens, Division of Natural Sciences,
Athens 10679, Greece}\\
\end{center}

\bigskip

\centerline{\bf ABSTRACT}

\noindent  
{\small The magnitude of B-mode polarization in the cosmic microwave background as
measured by BICEP2 favours models of chaotic inflation with a quadratic $m^2 \phi^2/2$
potential, whereas data from the Planck satellite favour a small value of the tensor-to-scalar
perturbation ratio $r$ that is highly consistent with the Starobinsky $R + R^2$ model.
Reality may lie somewhere between these two scenarios.
In this paper we propose a minimal two-field no-scale supergravity model that interpolates
between quadratic and Starobinsky-like inflation as limiting cases, while retaining the
successful prediction $n_s \simeq 0.96$.}

\vspace{0.2in}

\begin{flushleft}
April 2014
\end{flushleft}
\medskip
\noindent

\newpage

\section{Introduction}

The theory of inflationary cosmology has received important boosts
from the first release of data from the Planck satellite~\cite{Planck} - which confirmed
the infrared tilt of the scalar perturbation spectrum $n_s$~\cite{WMAP} expected in slow-roll
models of inflation~\cite{rev} - and the discovery by BICEP2 of B-mode
polarization fluctuations~\cite{BICEP2} - which may be interpreted as primordial
tensor perturbations with a large ratio $r$ relative to the scalar perturbations,
as would be generated in models with a large energy density during inflation.
On the other hand, whereas Planck and BICEP2 agree with earlier WMAP
data that $n_s \simeq 0.96$, there is tension between the BICEP2
measurement $r = 0.16^{+ 0.06}_{- 0.05}$ (after dust subtraction) and the Planck upper limit on $r$
obtained indirectly from the temperature fluctuations at low multipoles~\cite{tension}.
Theoretically, this tension could be reduced by postulating large running of the scalar
spectral index, but this would require a large deviation from the hitherto
successful slow-roll inflationary paradigm. Experimentally, it is known that
the temperature perturbations in the Planck data~\cite{Planck} lie somewhat below the
standard inflationary predictions at low multipoles, and there is a hint of
a hemispherical asymmetry, so perhaps the inflationary paradigm does
indeed require some modification~\cite{fix}. On the other hand, verification of the
dust model used by BICEP2 is desirable, and other possible sources of
foreground contamination such as magnetized dust associated with radio
loops need to be quantified~\cite{dust}. The good news is that the experimental situation should be
clarified soon, with additional data from Planck, the Keck Array and other
B-mode experiments.

In the mean time, inflationary theorists are having a field day exploring models
capable of accommodating the BICEP2 and/or Planck data. Taken at face
value, the BICEP2 data are highly consistent with the simplest possible
chaotic $m^2 \phi^2/2$ potential~\cite{m2} that predicts $r \simeq 0.16$, 
whereas the Planck data tend to favour the Starobinsky $R + R^2$ model~\cite{Staro,MC,Staro2}
that predicts $r \simeq 0.003$. It seems very likely that experimental measurements
of $r$ may settle down somewhere between these limiting cases, so it is
interesting to identify models that interpolate between them, while retaining
the successful prediction $n_s \simeq 0.96$. Desirable features of such models
would include characteristic predictions for other inflationary observables
and making connections with particle physics.

With the latter points in mind, we consider the natural framework for formulating
models of inflation to be supersymmetry~\cite{ENOT,nost,hrr,gl1,EENOS,enq}, specifically local supersymmetry, i.e.,
supergravity~\cite{SUGRA}. Moreover, in order to avoid holes in the effective potential
with depths that are ${\cal O}(1)$ in Planck units and to address the $\eta$
problem~\cite{eta}, we favour no-scale supergravity~\cite{no-scale,LN}, which has the theoretical advantage
that it emerges naturally as the low-energy effective field theory derived from
compactified string theory~\cite{Witten}. In the past we~\cite{ENO6,ENO7,ENO8,AHM,EGNO} and others
\cite{KLno-scale,WB,FKR,fklp,pallis,FeKR,oops} have shown how no-scale
supergravity with a Wess-Zumino or other superpotential~\cite{Cecotti} leads naturally to
Starobinsky-like inflation, and more recently we~\cite{EGNO} and others~\cite{gl2,bg,msy2,Davis:2008fv,Ant,abdk2,klr,klor,lln,bww} have given examples how
quadratic inflation may be embedded in no-scale supergravity.

In this paper we introduce a minimal two-field no-scale supergravity model with a
K\"ahler potential motivated by orbifold compactifications of string theory \cite{DKL,casas}.
The initial condition for the inflaton field has a free parameter that
which can be regarded as an angle in the two-field space. Varying this angle,
we can interpolate between the quadratic and Starobinsky-like limits: $0.16 \ga r \ga 0.003$~\footnote{One
might have thought that~\cite{FeKR} would provide a suitable framework for achieving this. However,
the Starobinsky-like limit is lost when the K\"ahler potential is stabilized as discussed in~\cite{EGNO,oops}.},
while $n_s \simeq 0.96$ for most values of the interpolating parameter. We follow numerically
the evolution of the inflaton in this space, including the possibility of initial values of the inflaton
fields that are larger than the minimal values need to obtain sufficient e-folds.
The key model predictions are
insensitive to the mechanism of supersymmetry breaking, as long as it occurs
at some scale much less than the inflaton mass. We illustrate this via a Polonyi
example of supersymmetry breaking, and comment on the connection to
particle phenomenology in this model.

The structure of this paper is as follows. In Section~2 we specify our
model, and in Section~3 we provide numerical analyses of inflationary
scenarios with different initial conditions for the inflaton field, discussing
its predictions for $n_s$ and  $r$. Then, in Section~4 we
discuss briefly supersymmetry breaking, and in Section~5 we draw some conclusions.

\section{Specification of the Model}

The original, minimal no-scale supergravity model has a K\"ahler potential
of the form~\cite{no-scale}
\beq
K\; = \; -3\ln(T+\bar{T})+\dots \, ,
\label{minimal}
\eeq
where the dots represent a possible superpotential, terms involving additional matter fields, etc..
Subsequent to its discovery, it was shown that no-scale supergravity emerges
naturally as the low-energy effective field theory in generic string compactifications~\cite{Witten}.
In general, these contain the complex moduli $T_i: i = 1, 2, 3$, and (\ref{minimal}) becomes
\beq
K \; = \; - \sum_{i=1}^3 \ln(T_i+\bar{T^i})+ \dots \, .
\label{3moduli}
\eeq
In the specific case of orbifold compactifications of string, matter fields $\phi$ have
non-zero modular weights and appear in a K\"ahler potential term of the form~\cite{DKL}
\beq
\Delta K \; = \; \frac{|\phi|^2}{\prod_{i=1}^3 (T_i+\bar{T^i})} \, .
\label{3weights}
\eeq
For simplicity, we consider here the case where the ratios of the three orbifold
moduli are fixed at a high scale by some unspecified mechanism so that,
neglecting irrelevant constants, the K\"ahler potential may be written in the form~\cite{casas}:
\beq
K \; = \; -3\ln(T+\bar{T})+\frac{|\phi|^2}{(T+\bar{T})^3} \, .
\label{ourK}
\eeq
Thus we assume that the field $\phi$ has modular weight $3$,
and stress that the special properties of the model we describe below depend on this
choice of modular weight.
More general constructions of Minkowski and deSitter vacua were discussed
in a similar context in~\cite{lnr}, which focused on models where the modular weight of $\phi$ is 0.

The specification of our no-scale supergravity model of inflation is completed
by specifying the superpotential:
\beq
W \; = \; \sqrt{\frac{3}{4}}\,\frac{m}{a}\phi(T-a) \, ,
\label{ourW}
\eeq
and we identify $T$ as the chiral (two-component) inflaton superfield.
Although the superpotential (\ref{ourW}) is the same as in the model \cite{Cecotti}
derived in a realization of $R + R^2$ gravity in a SU(2,1)/SU(2)$\times$U(1) no-scale model,
the \K\ potential (\ref{ourK}) manifests only the minimal SU(1,1)/U(1) no-scale structure.
The scalar potential is derived from the \K\ potential (\ref{ourK}) and the superpotential (\ref{ourW})  through
\beq
V=e^{ K}\left({ K}^{i\bar j}D_i W\bar D_{\bar j}\bar{ W}-3| W|^2\right),
\label{eqn:SUGRApotential}
\eeq
where $D_i W\equiv\partial_i W+{ K}_iW$.
We will work in Planck units $M_P^2 = 8\pi G_N=1$.
The motion of the scalar field $\phi$ is constrained by the exponential factor $e^K$:
\beq
V\; \propto \; e^{|\phi|^2/(T+\bar{T})^3} \; \simeq \; e^{(2a)^{-3}|\phi|^2} \, .
\label{fixphi}
\eeq
For $a\leq1$, it can be assumed (and we have verified) that $\phi$ is rapidly driven to zero at
the onset of inflation, and we assume that $a=1/2$ so that $\phi$ has a canonical kinetic term. 
For vanishing $\phi$, 
the scalar potential takes the simple form
\beq
V = \frac{3 m^2}{4 a^2} |T-a|^2
\eeq
which can be shown to be equivalent to the Starobinsky model
along the real direction of the canonical field associated with $T$ \cite{Cecotti,KLno-scale,ENO7}.
For other choices of the modular weight $w$ of $\phi$,
the potential (at $\phi = 0$)
is
\beq
V = \frac{3 m^2}{4 a^2} |T-a|^2 (T+T^*)^{(w-3)} \, .
\eeq
It is convenient to decompose $T$ into its real and imaginary parts defined by
$\rho$ and $\sigma$, respectively, where $\rho$ is canonical and $\sigma$ is canonical at the minimum
when $\rho = 0$:
\beq
T \; = \; a\left(e^{-\sqrt{\frac{2}{3}}\rho}+i\sqrt{\frac{2}{3}}\,\sigma\right)
\label{rhosigma}
\eeq
The scalar component of $T$ minimizes the potential when $T=a$, and
the resulting Lagrangian
is given by
\begin{eqnarray}
\mathcal{L} \; & =  & \; \frac{1}{2}\partial_{\mu}\rho\partial^{\mu}\rho+\frac{1}{2}e^{2\sqrt{\frac{2}{3}}\rho}\partial_{\mu}\sigma\partial^{\mu}\sigma \nonumber \\
& & - \; \frac{3}{2^{(5-w)}a^{(3-w)}}m^2e^{\sqrt{\frac{2}{3}}(3-w)\rho}\left(1-e^{-\sqrt{\frac{2}{3}}\rho} \right)^2 
\nonumber \\
 & & - \; \frac{1}{2^{(4-w)}a^{(3-w)}}m^2e^{\sqrt{\frac{2}{3}}(3-w)\rho}\sigma^2 \, .
\label{effLw}
\end{eqnarray}
In the particular case $w = 3$ this reduces to 
\beq
\mathcal{L} \; = \; \frac{1}{2}\partial_{\mu}\rho\partial^{\mu}\rho+\frac{1}{2}e^{2\sqrt{\frac{2}{3}}\rho}\partial_{\mu}\sigma\partial^{\mu}\sigma - 
\frac{3}{4}m^2\left(1-e^{-\sqrt{\frac{2}{3}}\rho} \right)^2 - \frac{1}{2}m^2\sigma^2 \, ,
\label{effL}
\eeq
which is the starting-point for our analysis. 

Although this is similar to the Lagrangian for the no-scale model
of~\cite{Cecotti,FeKR}, it differs in an important way. In that SU(2,1)/SU(2)$\times$U(1) model,
the mass term for $\sigma$ contained a coupling to $\rho$ of the form $e^{2\sqrt{2/3}\rho}$.
In (\ref{effL}), the real and imaginary parts of $T$ are decoupled
in the potential and only mix through their kinetic terms (we return later to the effect of this mixing).

The minimum of the effective potential (\ref{effL}) in the $(\rho,\sigma)$ plane is located at
\beq
\rho_0 \; = \; \sigma_0 \; = \; 0 \, .
\label{minimum}
\eeq
When $\rho$ is at the minimum, the effective Lagrangian for $\sigma$ is
\beq
\mathcal{L}=\frac{1}{2}\partial_{\mu}\sigma\partial^{\mu}\sigma -  \frac{1}{2}m^2\sigma^2 \, ,
\label{Lsigma}
\eeq
and we recover the minimal quadratic inflationary model. Conversely, 
when $\sigma$ is at the minimum, the effective Lagrangian for $\rho$ is
\beq
\mathcal{L} \; = \; \frac{1}{2}\partial_{\mu}\rho\partial^{\mu}\rho - 
\frac{3}{4}m^2\left(1-e^{-\sqrt{\frac{2}{3}}\rho} \right)^2 \, ,
\label{Lrho}
\eeq
which is of the same form as the Starobinsky model~\cite{Staro}.

\begin{figure}[h!]
\centering
	\scalebox{0.55}{\includegraphics{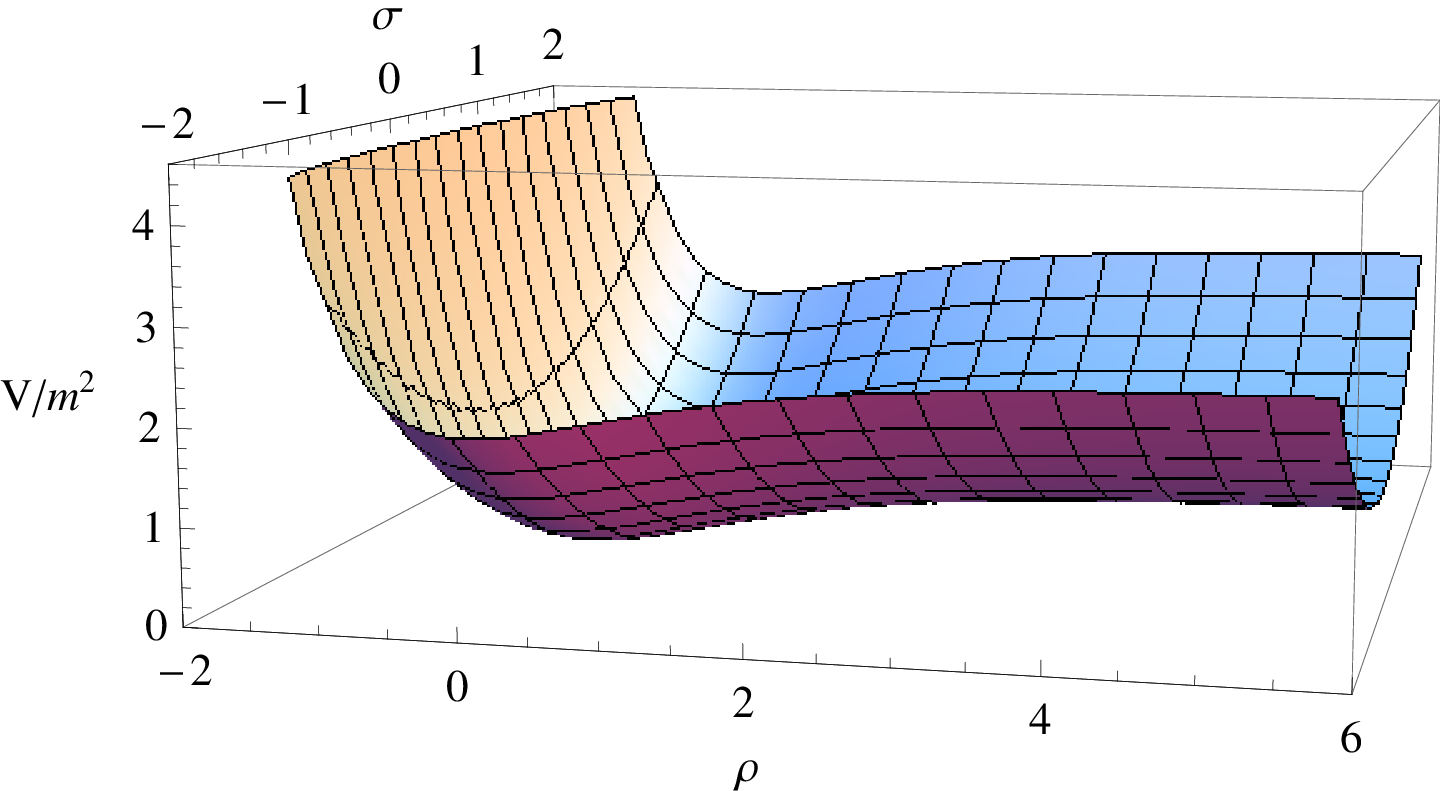}} 
	\scalebox{0.45}{\includegraphics{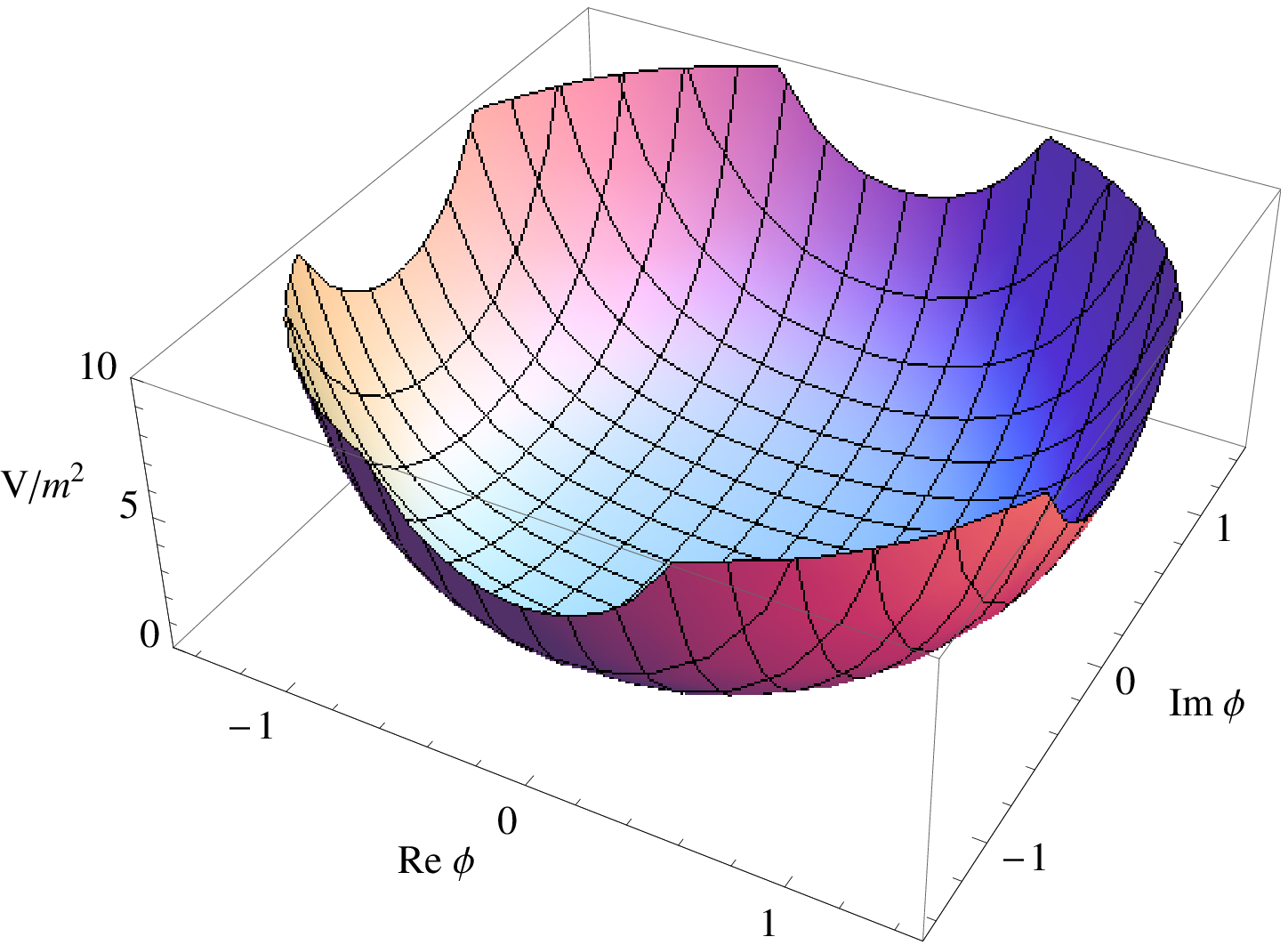}} \\
	\scalebox{0.49}{\includegraphics{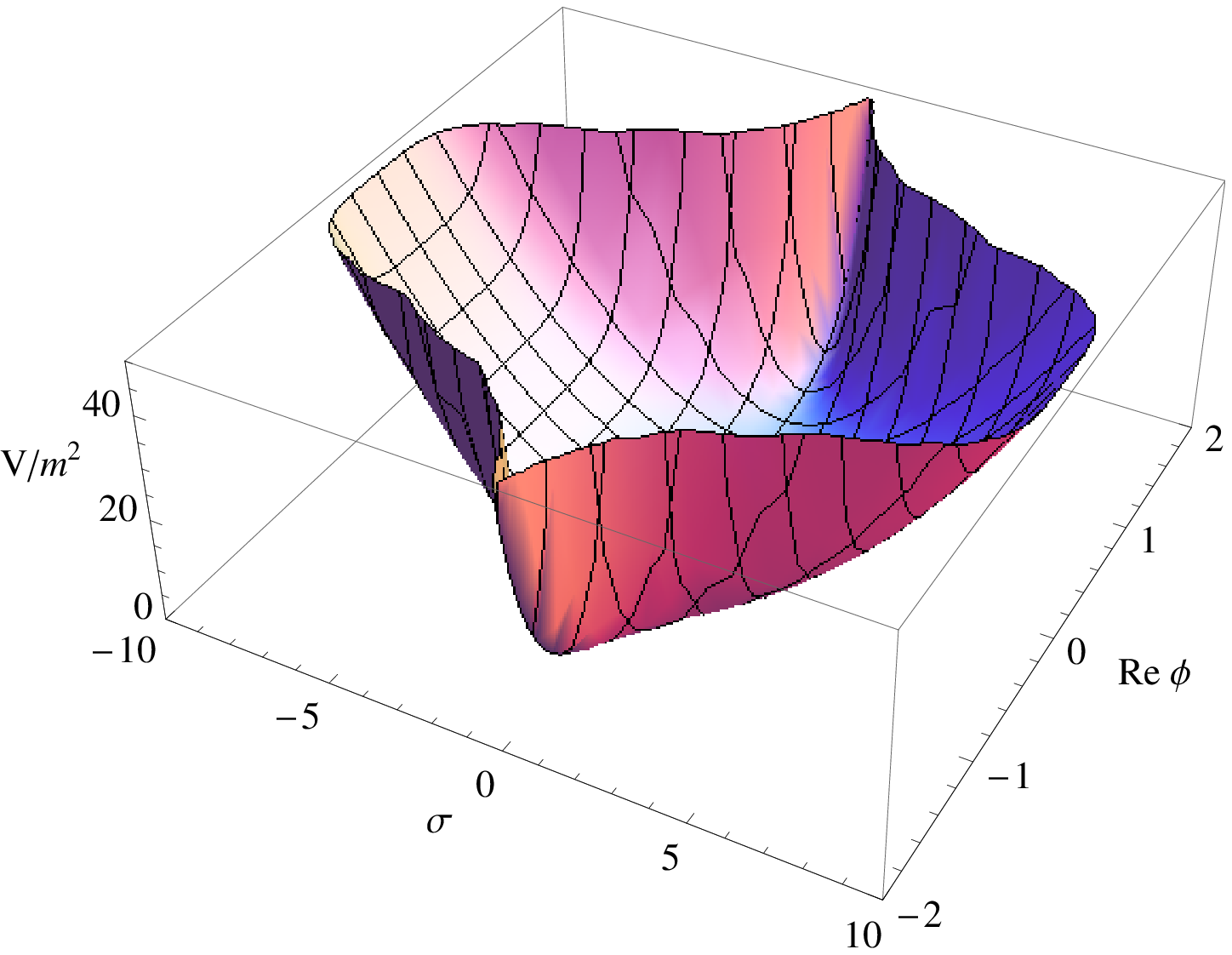}} 
	\scalebox{0.49}{\includegraphics{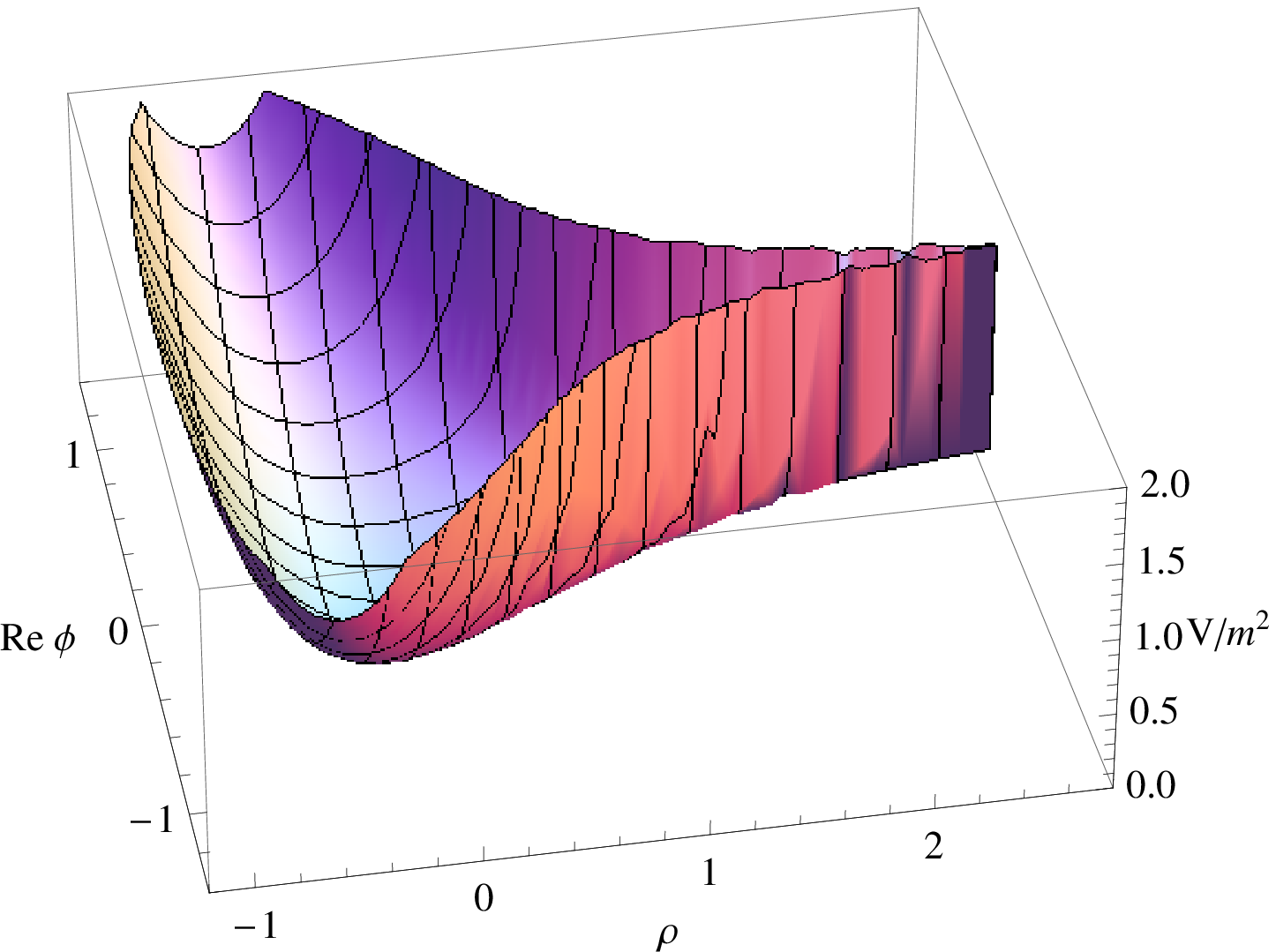}} 
	\caption{\it Slices through the effective potential for the model (\protect\ref{effL}).
	Upper left: The potential in the $(\rho, \sigma)$ plane for $\phi = 0$.
	Upper right: The potential in the $({\rm Re} \, \phi, {\rm Im} \, \phi)$ plane for $\rho=\sigma = 0$.
	Lower left: The potential in the $(\sigma, {\rm Re} \, \phi)$ plane for $\rho={\rm Im} \, \phi=0$.
	Lower right: The potential in the $(\rho, {\rm Re} \, \phi)$ plane for $\sigma={\rm Im} \, \phi=0$.} 
	\label{fig:potpix}
\end{figure}

Various slices through the potential (\ref{effL}) for the canonical choice $a=1/2$
are shown in Fig.~\ref{fig:potpix}. In the upper left panel we see the characteristic
Starobinsky form in the $\rho$ direction (\ref{Lrho}) and the simple quadratic form in the
$\sigma$ direction (\ref{Lsigma}). In the upper right panel we see that both the real
and imaginary parts of $\phi$ are indeed stabilized at zero, the value that was assumed in the
upper left panel. The lower left panel shows that the effective potential is well-behaved in the
$(\sigma, {\rm Re} \, \phi)$ plane  for $\rho={\rm Im} \, \phi=0$, and the lower right panel
makes the same point for the $(\rho, {\rm Re} \, \phi)$ plane for $\sigma={\rm Im} \, \phi=0$.
In both cases, the effective potential is identical when the r\^oles of ${\rm Re} \, \phi$
and ${\rm Im} \, \phi$ are reversed. The Starobinsky form of potential is visible again in the
lower right panel of Fig.~\ref{fig:potpix}. 

\begin{figure}[!h]
\centering
	\scalebox{0.8}{\includegraphics{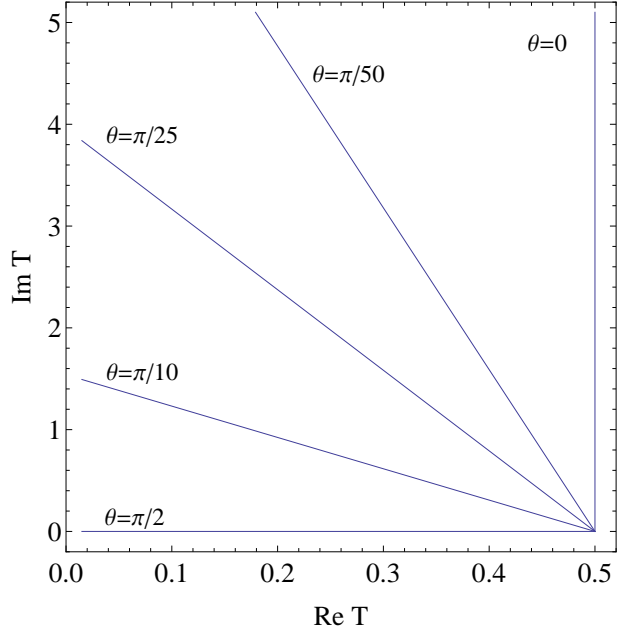}} 
	\caption{\it Inflationary directions in the $({\rm Re}\,T, {\rm Im}\,T)$ plane, labeled by the stabilization angle $\theta$.} 
	\label{fig:Tplane0}
\end{figure}

The model described by (\ref{ourK}) and (\ref{ourW}) has two dynamical fields, and a correct discussion of their behaviour during inflation requires a more sophisticated analysis than single-field models of inflation \cite{2field,Turzynski}. We leave
such a discussion for future work~\cite{EGNO3}. Instead, here we modify the K\"ahler potential (\ref{ourK}) so as to reduce it to a family of nearly single-field models characterized by an angle $\theta$ in the $({\rm Re}\,T, {\rm Im}\,T)$ plane defined in Fig.~\ref{fig:Tplane0}. This is accomplished by introducing a $\theta$-dependent stabilization term of the same general form as introduced in~\cite{EKN}:
\begin{equation}
K \; = \; -3\log\left(T+T^* - c( \cos \theta(T+T^*-1) - i \sin \theta(T-T^*) )^4 \right) + \frac{|\phi|^2}{(T+T^*)^3} \, .
\label{stabilize}
\end{equation}
It is clear that, for a large enough coefficient $c$ of the quartic stabilization term, the inflaton
trajectory is confined to a narrow valley in field space, much like a bobsleigh confined inside
a narrow track.

\section{Numerical Analysis of the Model}


The classical motion of the inflaton field for the model (\ref{ourW}, \ref{stabilize}) can be numerically calculated solving the equations
\begin{align}
H^2 \; = \; \frac{1}{3}\left[ K_{a\bar{b}}\dot{\Psi}^a\dot{\bar{\Psi}}^{\bar{b}} + V(\boldsymbol{\Psi}) \right] &= \; (\dot{N})^2\ ,\\
\ddot{\Psi}^a + 3H\dot{\Psi}^a + \Gamma^a_{bc}\dot{\Psi}^b\dot{\Psi}^c + K^{a\bar{b}}\frac{\partial V}{\partial \bar{\Psi}^{\bar{b}}} &= \; 0\ ,
\end{align}
where $\boldsymbol{\Psi} \equiv (T,\phi)$, $K_{a\bar{b}}$ is the K\"ahler metric, $\Gamma^a_{bc} \equiv K^{a\bar{d}}\partial_{b}K_{c\bar{d}}$,  and $N$ is the number of e-foldings. The figures that follow show the resulting evolution for the $T$ and $\phi$ fields for four different choices of initial conditions. 

In order to calculate the values of the scalar tilt $n_s$ and the tensor-to-scalar ratio $r$ one cannot use the usual single field formulae, since isocurvature perturbations are generally present when more than one scalar field evolves during inflation. In order to simplify the analysis, we will calculate $n_s$ and $r$ assuming that $\phi$ starts at zero or is driven very quickly to the origin
using the techniques in~\cite{2field}.
The scalar tilt and the tensor to scalar ratio are then calculated from their definitions
\beq
n_s=1+\frac{d\log \mathcal{P}_{\mathcal{R}}}{d\log k}\ , \ \ r=\frac{\mathcal{P}_{T}}{\mathcal{P}_{\mathcal{R}}}\ .
\eeq
where $\mathcal{P}_{\mathcal{R}}$ is the power spectrum of the adiabatic perturbations.
The tensor perturbations have the same form as in the single field case, $\mathcal{P}_{T}=\left.\frac{2}{\pi^2}H^2\right|_{k=aH}$, since at linear order the scalar field perturbations decouple from vector and tensor perturbations \cite{perts}. For the scale $k$ we choose a perturbation that leaves the horizon at the start of the last 50 or 60 e-foldings of inflation, assuming that its value corresponds to the Minkowski-like vacuum 10 e-foldings before the scale leaves the Hubble radius.

\begin{figure}[h!]
\centering
	\scalebox{0.62}{\includegraphics{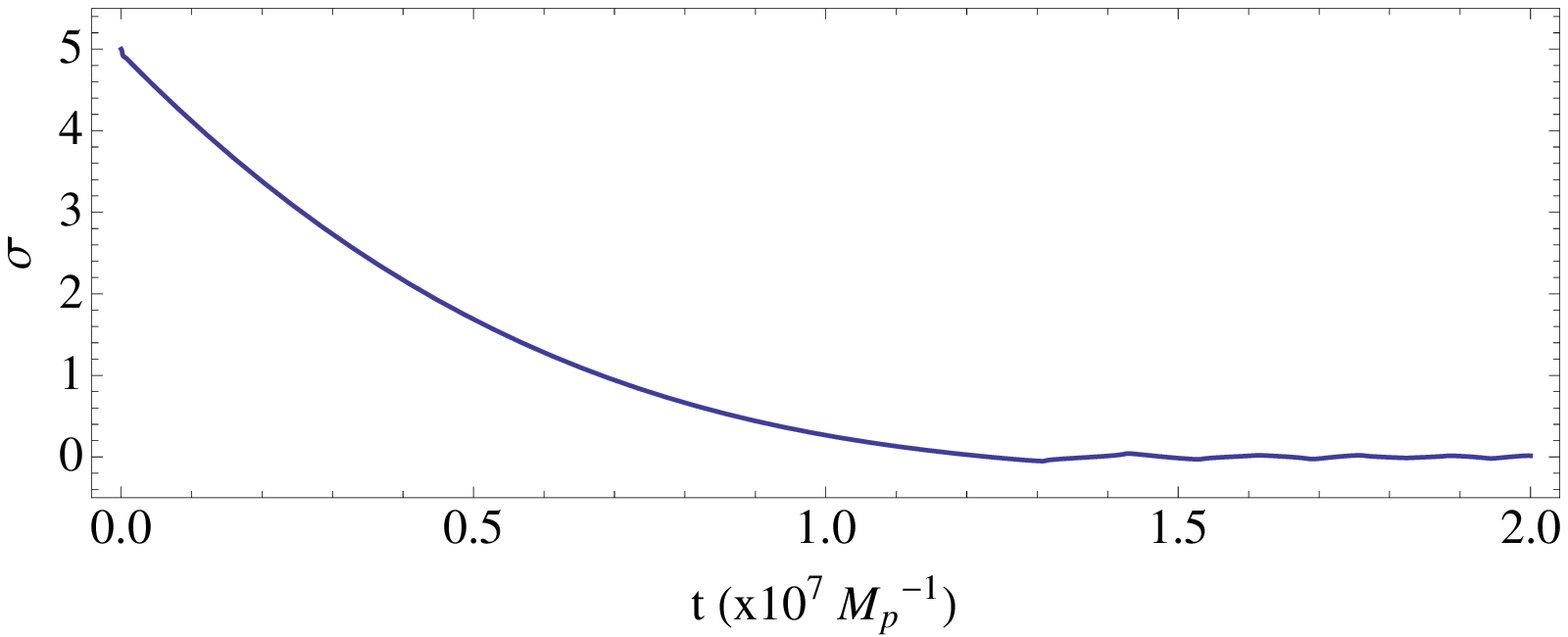}}
	\scalebox{0.62}{\includegraphics{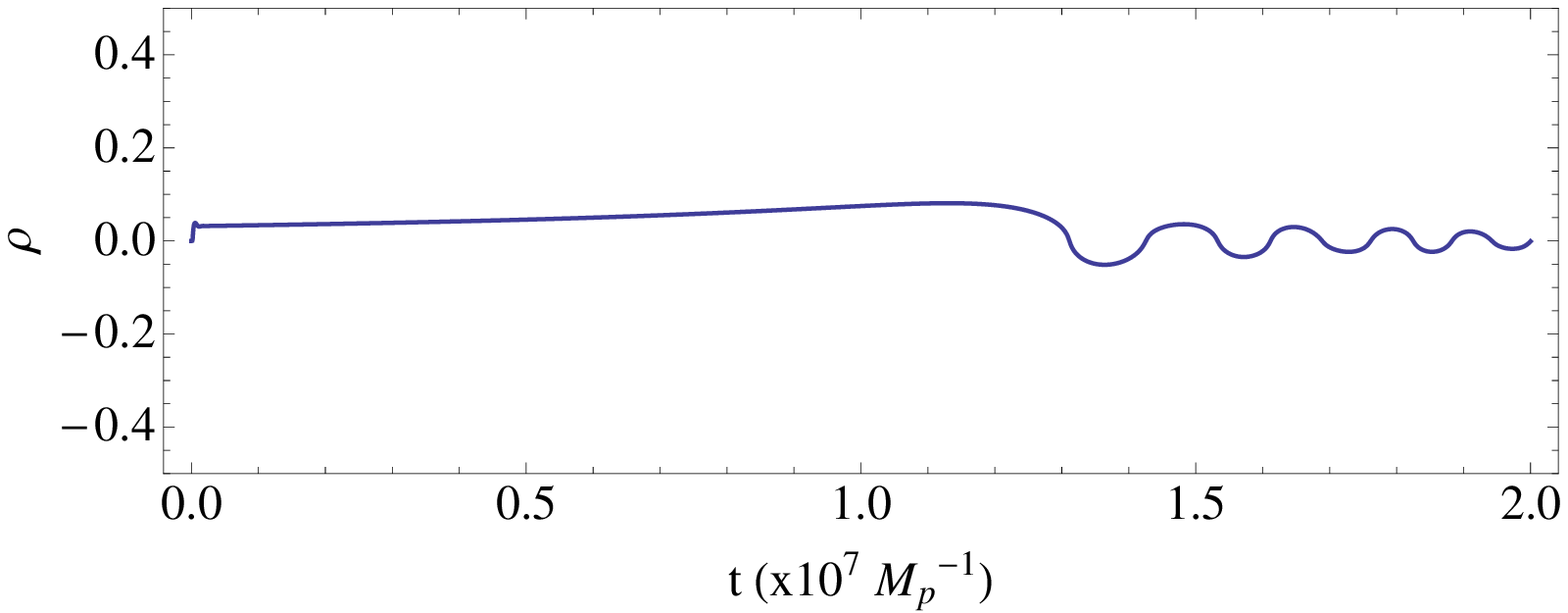}} 
	\scalebox{0.62}{\includegraphics{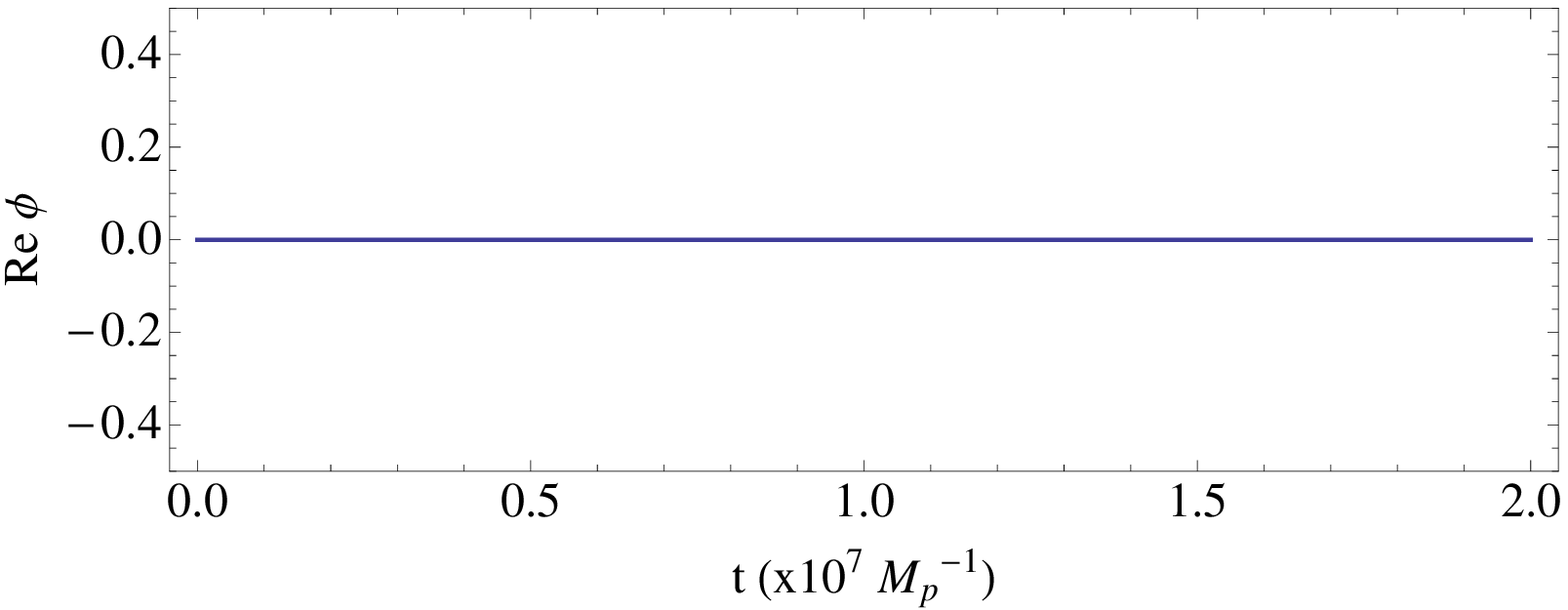}} 
	\scalebox{0.62}{\includegraphics{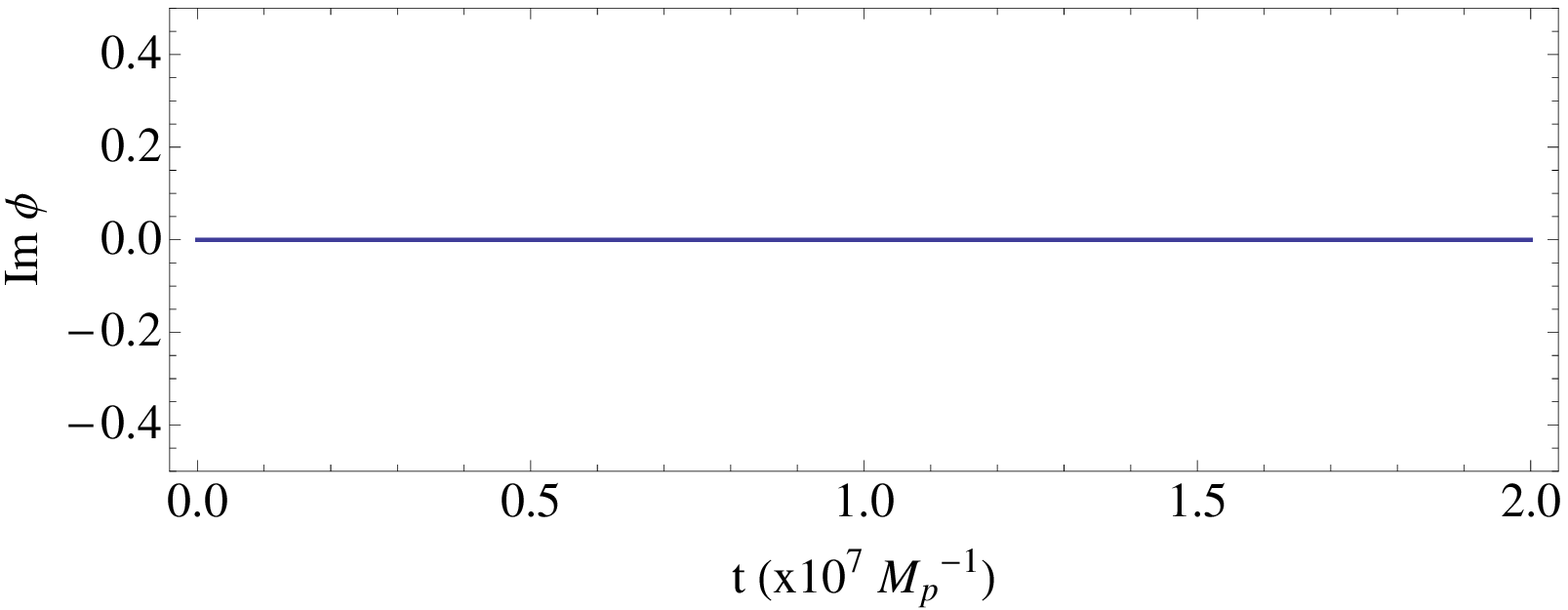}} 
	\scalebox{0.62}{\includegraphics{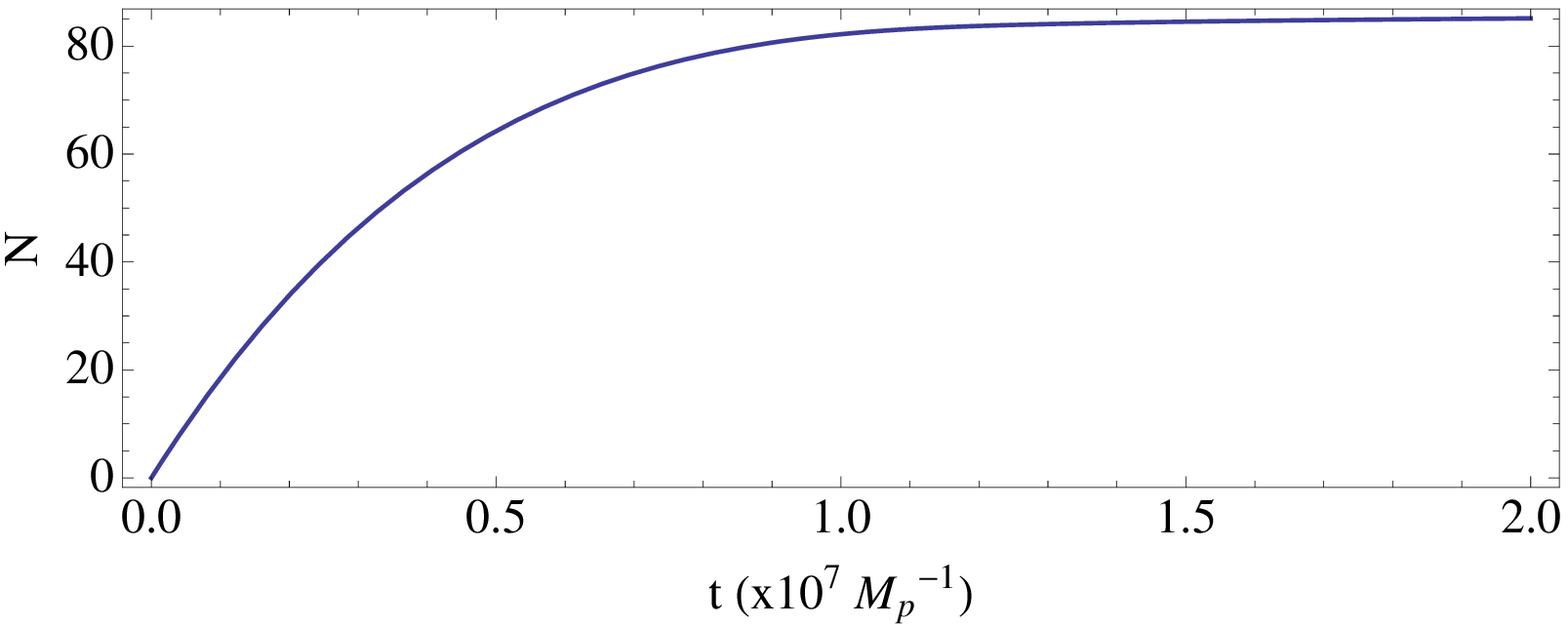}} 
	%
	\caption{\it Numerical solution for the choice $\theta=0$, $c = 1000$, with initial conditions $\rho_0=0$, $\sigma_0=5$ and $\phi_0=0$:
	field evolution and e-folds.} 
	\label{fig:sol11}
\end{figure}

\begin{figure}[!h]
\centering
	\vspace{15pt}
	\scalebox{0.9}{\includegraphics{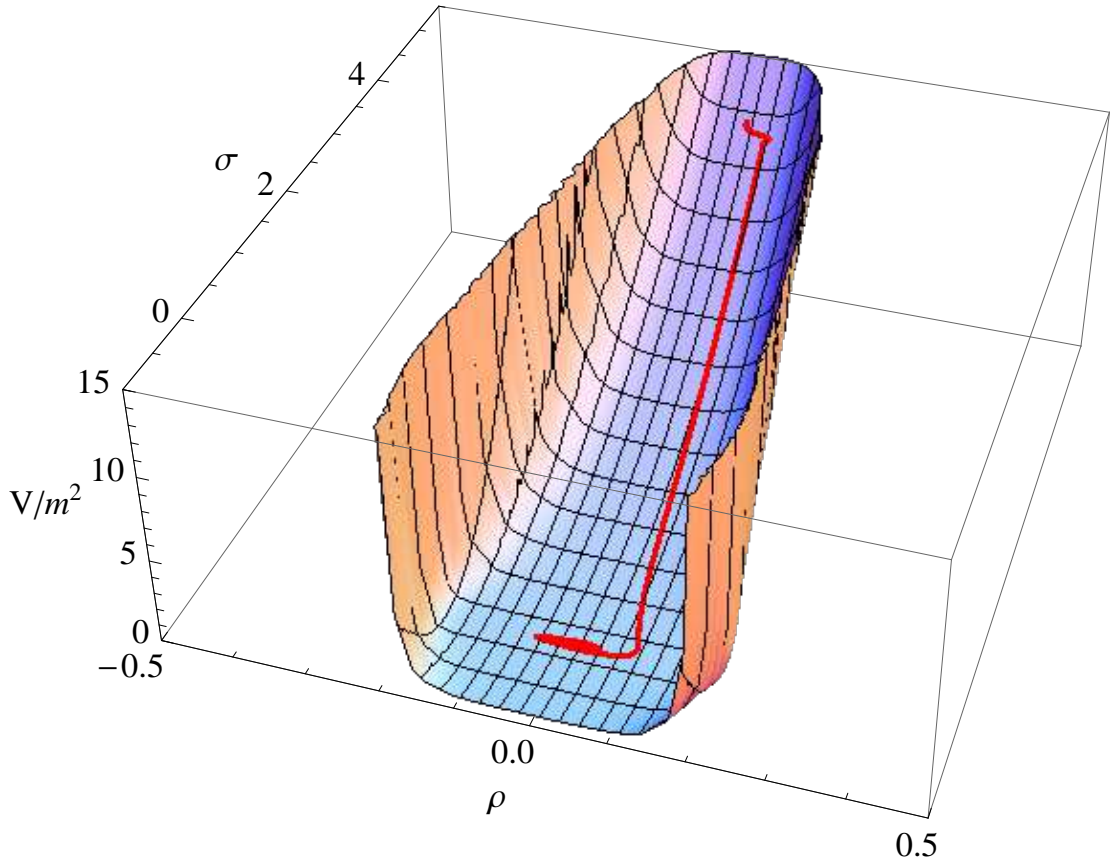}} 
	\caption{\it Numerical solution for $\theta=0$, $c = 1000$, with initial conditions $\rho_0=0$, $\sigma_0=5$ and $\phi_0=0$:
	`circling the drain'.}
	\label{fig:sol12}
\end{figure}

Figs.~\ref{fig:sol11} and \ref{fig:sol12} display the numerical solution for $\theta=0$ and $c = 1000$~\footnote{As discussed
later, we have checked that
these and subsequent results are insensitive to the value of $c$.}, with the initial
conditions $\rho_0=0$, $\sigma_0=5$ and $\phi_0=0$ , corresponding to the case of quadratic inflation.
The top panel of Fig.~\ref{fig:sol11} shows the evolution of $\sigma$, which is the inflaton field in this case.
The second panel of Fig.~\ref{fig:sol11} displays the evolution of $\rho$. 
We note a perturbation of $\rho$ that is due to its coupling with $\sigma$ through the kinetic term (see (\ref{effL})).
However, the value of $\rho$ remains small and does not
affect substantially the inflationary dynamics of the inflaton field other than allowing for
more e-folds of inflation at smaller values of $\sigma$ (which is not
canonically normalized when $\rho \ne 0$).  The third panel shows that
${\rm Re} \, \phi$ remains zero (and the same is true for ${\rm Im} \, \phi$). 
Finally, the bottom panel of Fig.~\ref{fig:sol11} displays the growth of the number of e-folds for this choice
of boundary conditions. We find the following values of $n_s$ and $r$ for this case:
\begin{eqnarray}
N \; = \; 50: \; \; (n_s, r) & =  & (0.951, 0.088) \, , \nonumber \\
N \; = \; 60: \; \; (n_s, r) & = & (0.959, 0.074) \, .
\label{nsr1}
\end{eqnarray}
As already commented, small values of $\rho$ are generated during the evolution of the
inflaton field, and Fig.~\ref{fig:sol12} displays the joint evolution of $\rho$ and $\sigma$,
where we see a `circling the drain' phenomenon towards the end of inflation. This has the effect of reducing the value of $r$ by a factor of 2.6, without having a significant effect on $n_s$. A smaller reduction factor, and hence a larger value of $r \sim 0.16$,
is possible with a different stabilization term in the K\"ahler potential~\cite{EGNO3}.

\begin{figure}[!h]
\centering
	\scalebox{0.63}{\includegraphics{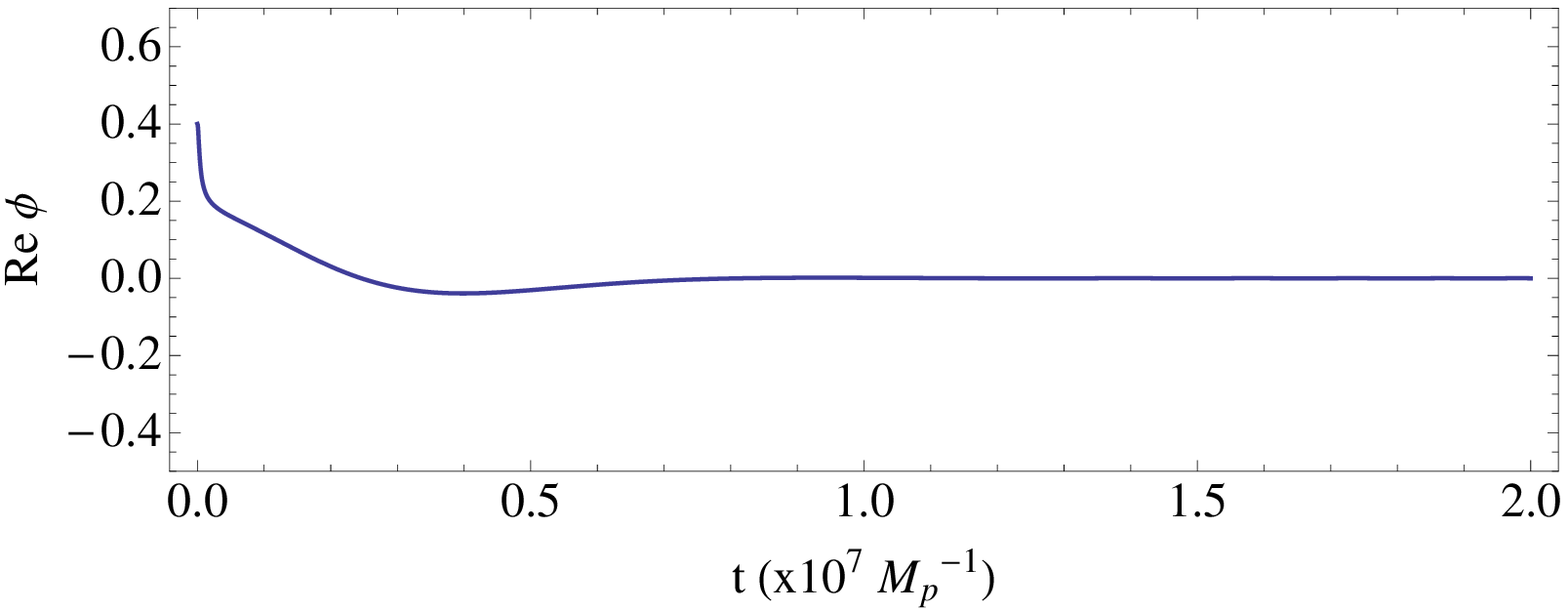}} 
	\scalebox{0.63}{\includegraphics{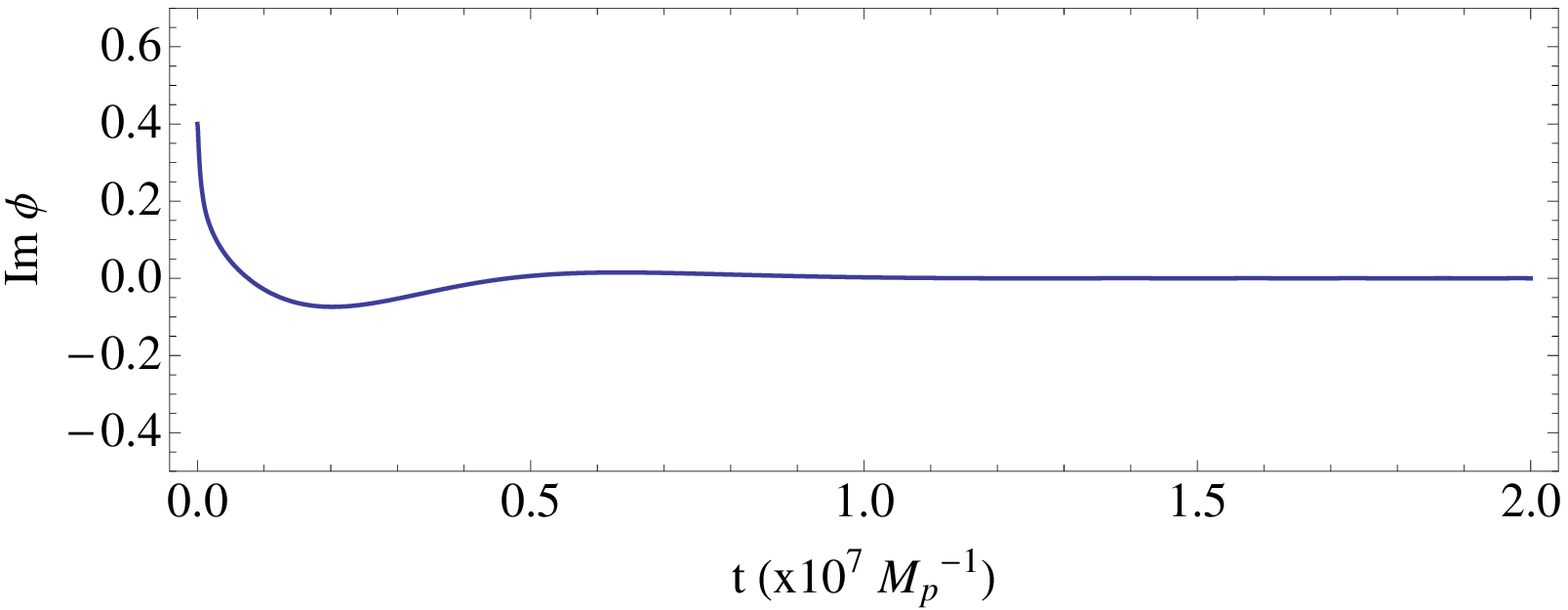}} 
	\scalebox{0.63}{\includegraphics{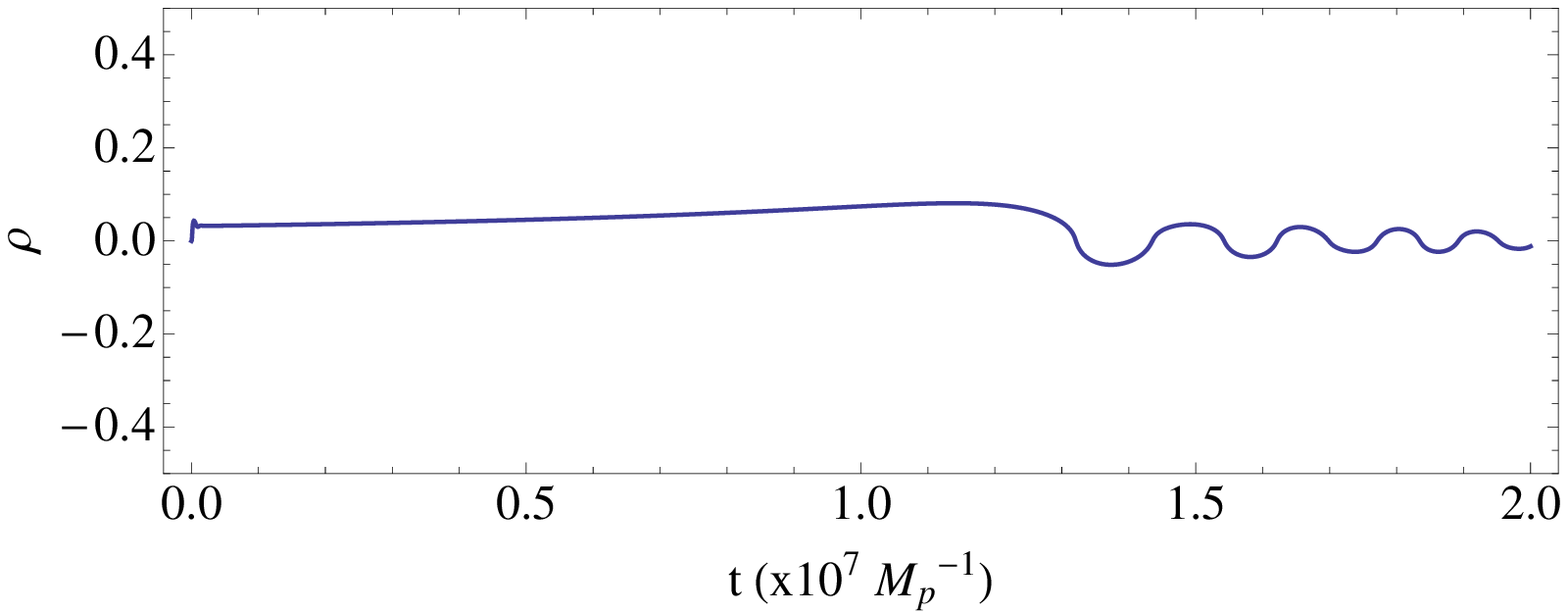}} 
	\scalebox{0.63}{\includegraphics{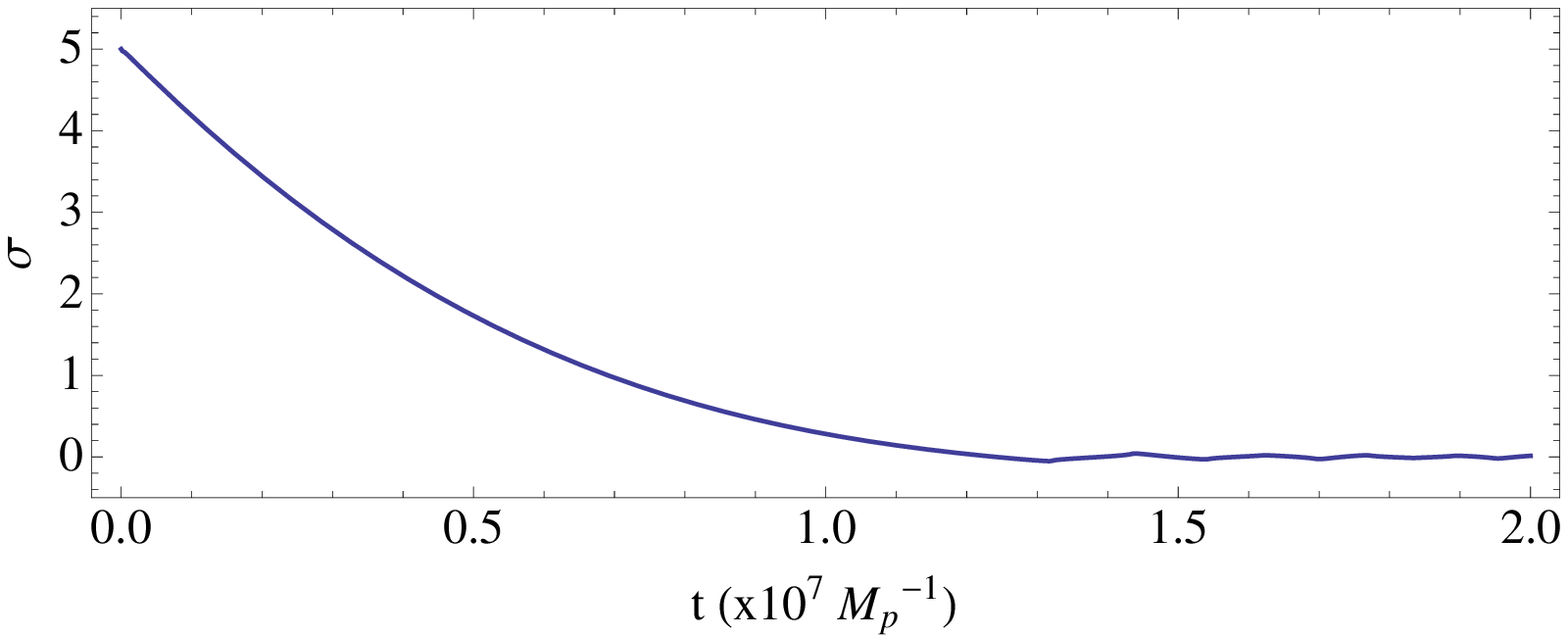}} 
	\scalebox{0.63}{\includegraphics{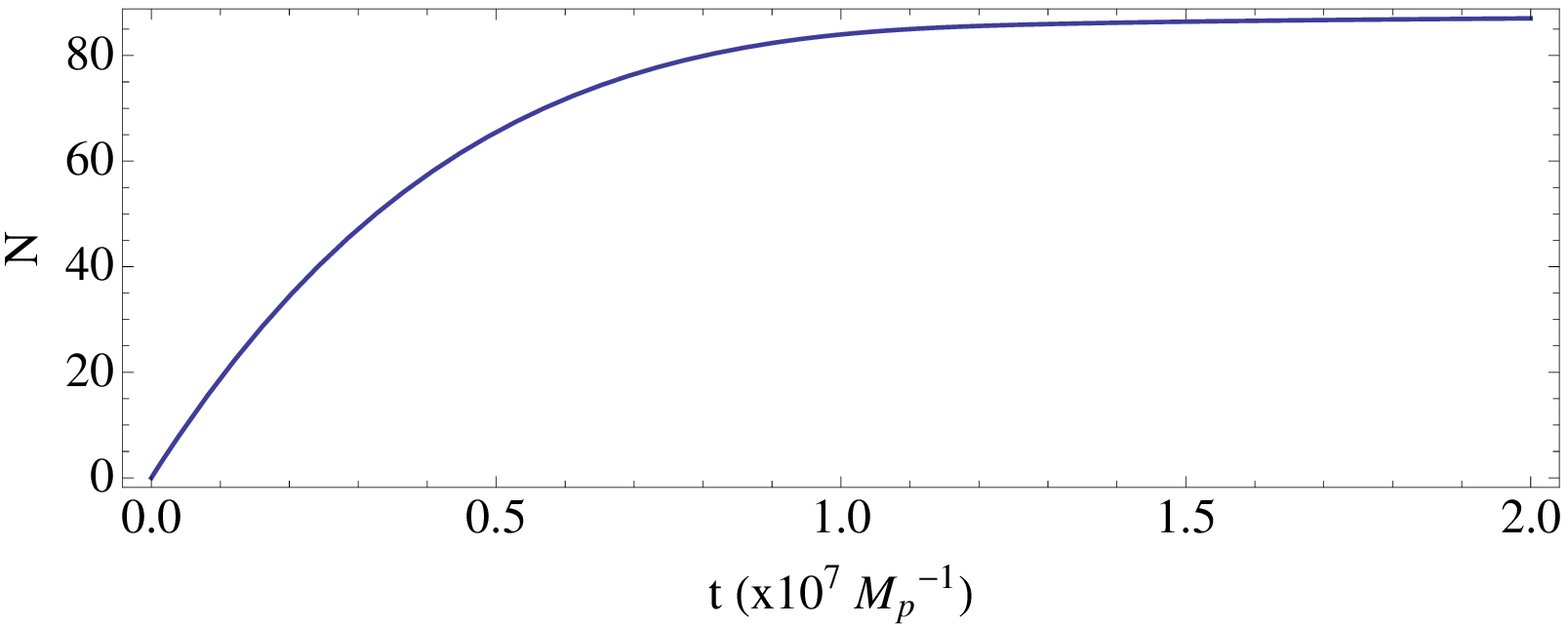}} 
	\caption{\it Numerical solution for $\theta=0$, $c = 1000$, with initial conditions $\rho_0=0$, $\sigma_0=5$ and $\phi_0=0.4+0.4i$.}
	 \label{fig:sol2}
\end{figure}

Fig.~\ref{fig:sol2} displays the numerical solution for $\theta=0$ and $c = 1000$, with the initial
conditions $\rho_0=0$, $\sigma_0=5$ and $\phi_0=0.4+0.4i$, corresponding to
a modification of the previous case of quadratic inflation, allowing for non-trivial evolution of $\phi$.
However, we see in the top panels of Fig.~\ref{fig:sol2} that both ${\rm Re} \, \phi$ and ${\rm Im} \, \phi$ evolve
rapidly to zero, as expected, and $\rho$ behaves similarly to the first case above (third panel).
Correspondingly, the behaviour of the inflaton $\sigma$ is also similar (fourth panel), and the
number of e-folds grows in a similar way as in the first case. We find that
\begin{eqnarray}
N \; = \; 50: \; \; (n_s, r) & =  & (0.943,0.077) \, , \nonumber \\
N \; = \; 60: \; \; (n_s, r) & = & (0.945, 0.058) \, ,
\label{nsr2}
\end{eqnarray}
results that are very similar to the first case.

\begin{figure}[!h]
\centering
	\scalebox{0.6}{\includegraphics{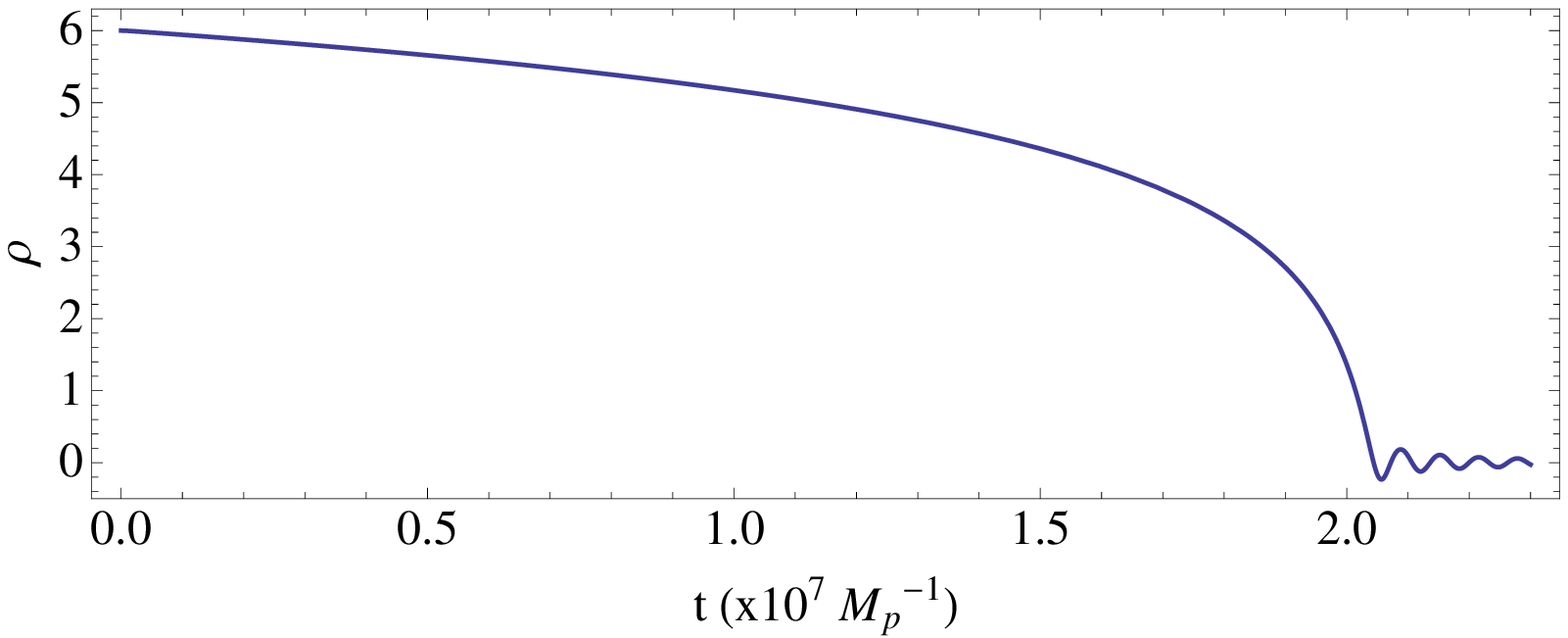}} 
	\scalebox{0.63}{\includegraphics{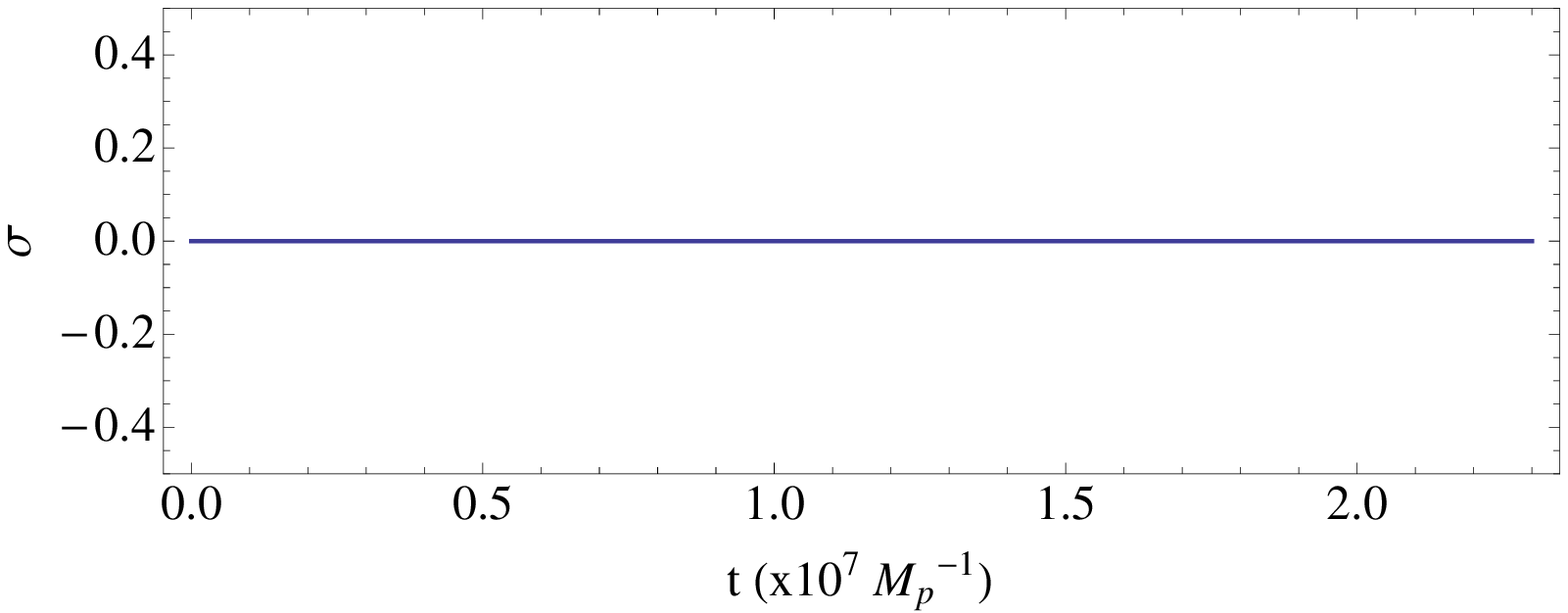}} 
	\scalebox{0.63}{\includegraphics{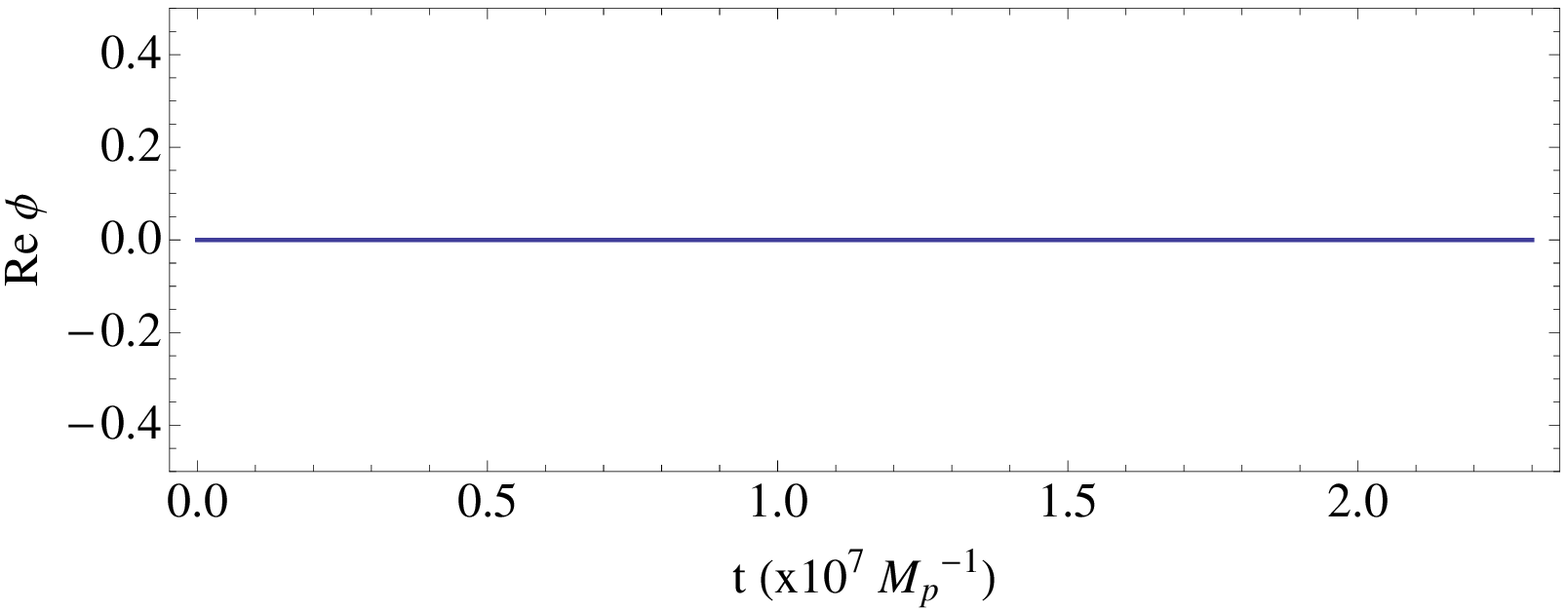}} 
	\scalebox{0.63}{\includegraphics{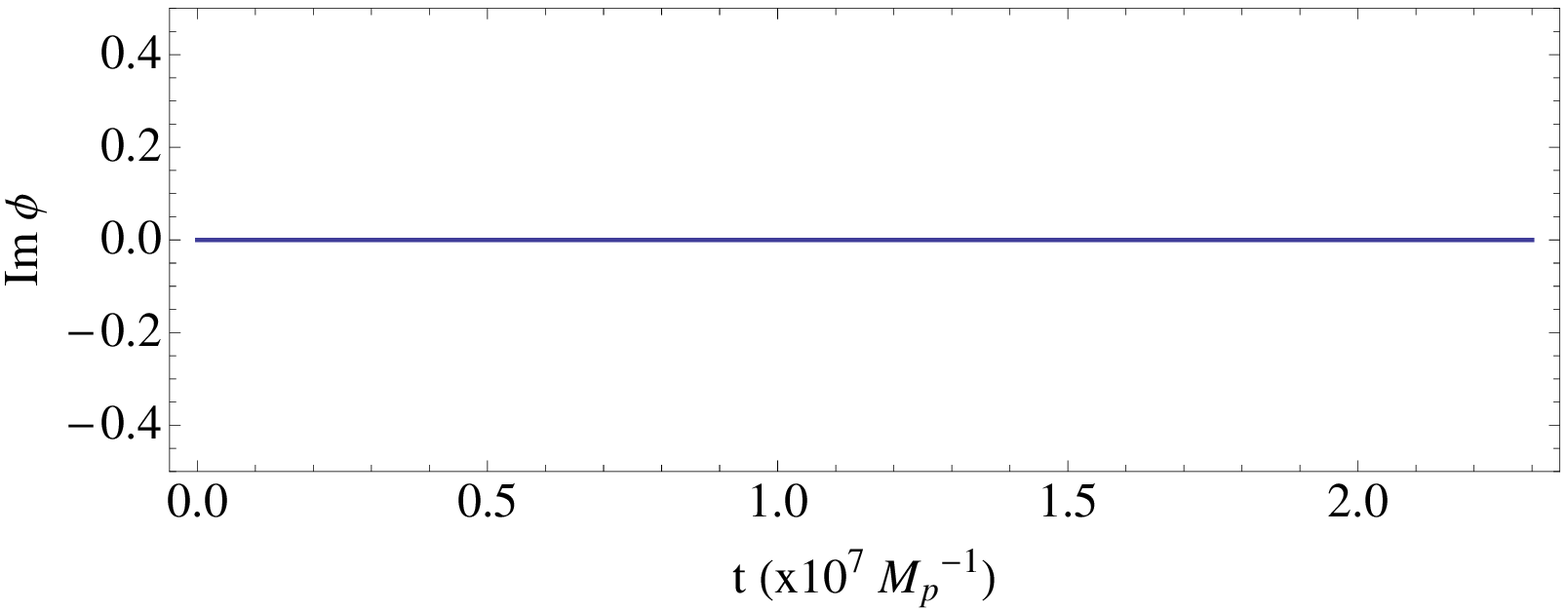}} 
	\scalebox{0.63}{\includegraphics{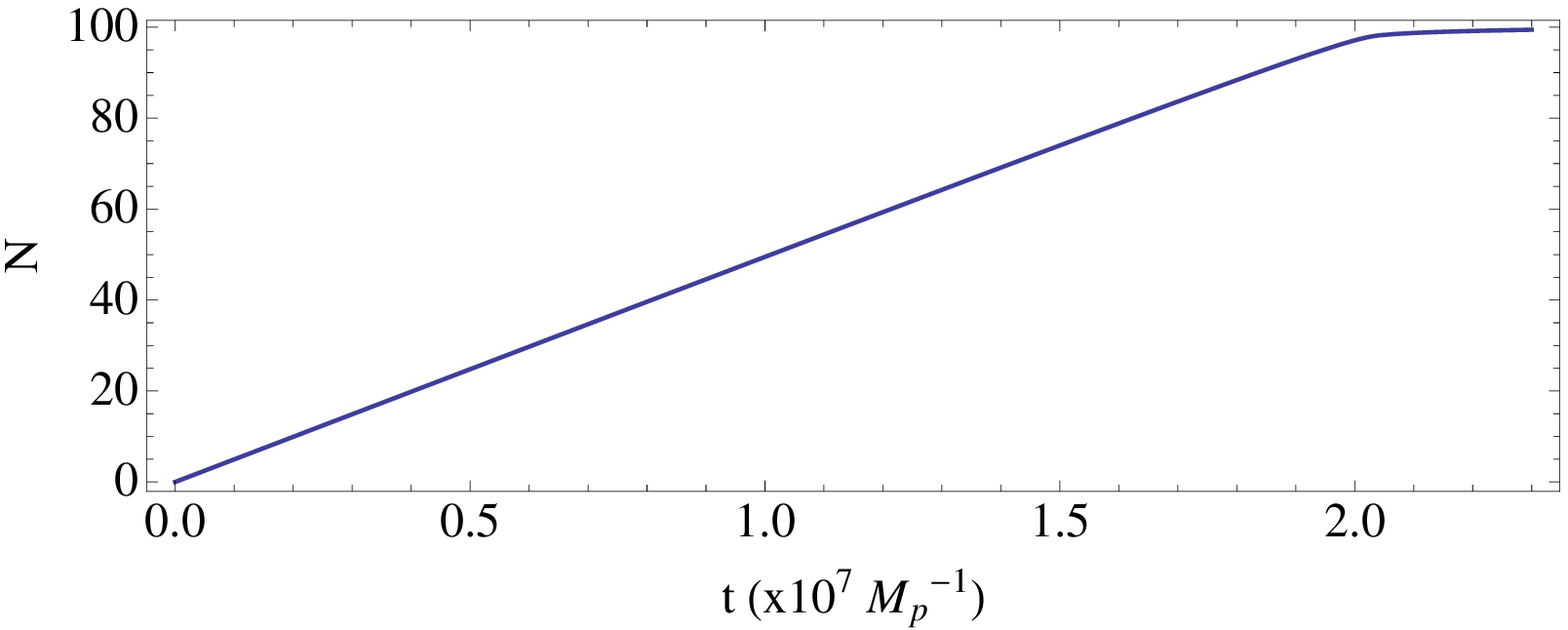}} 
	\caption{\it Numerical solution for $\theta=\pi/2$, $c = 1000$, with initial conditions $\rho_0=6$, $\sigma_0=0$ and $\phi_0=0$.} 
	\label{fig:sol3}
\end{figure}

\begin{figure}[!h]
\centering
	\scalebox{0.63}{\includegraphics{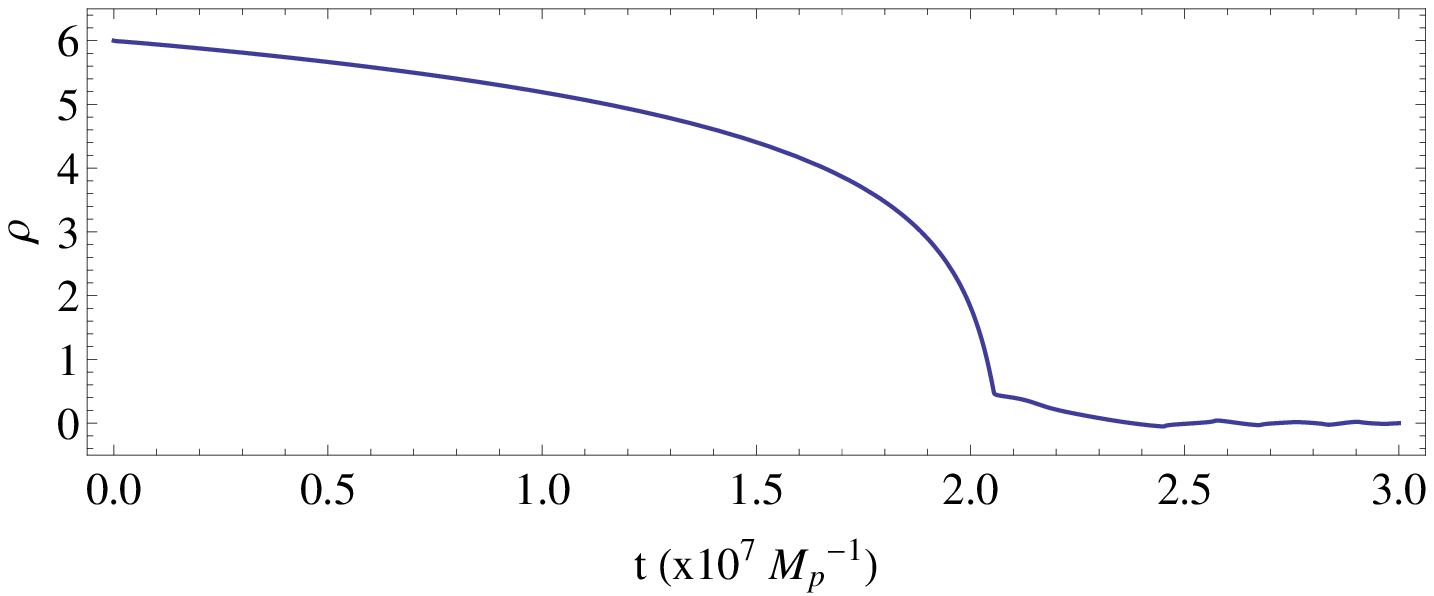}} 
	\scalebox{0.63}{\includegraphics{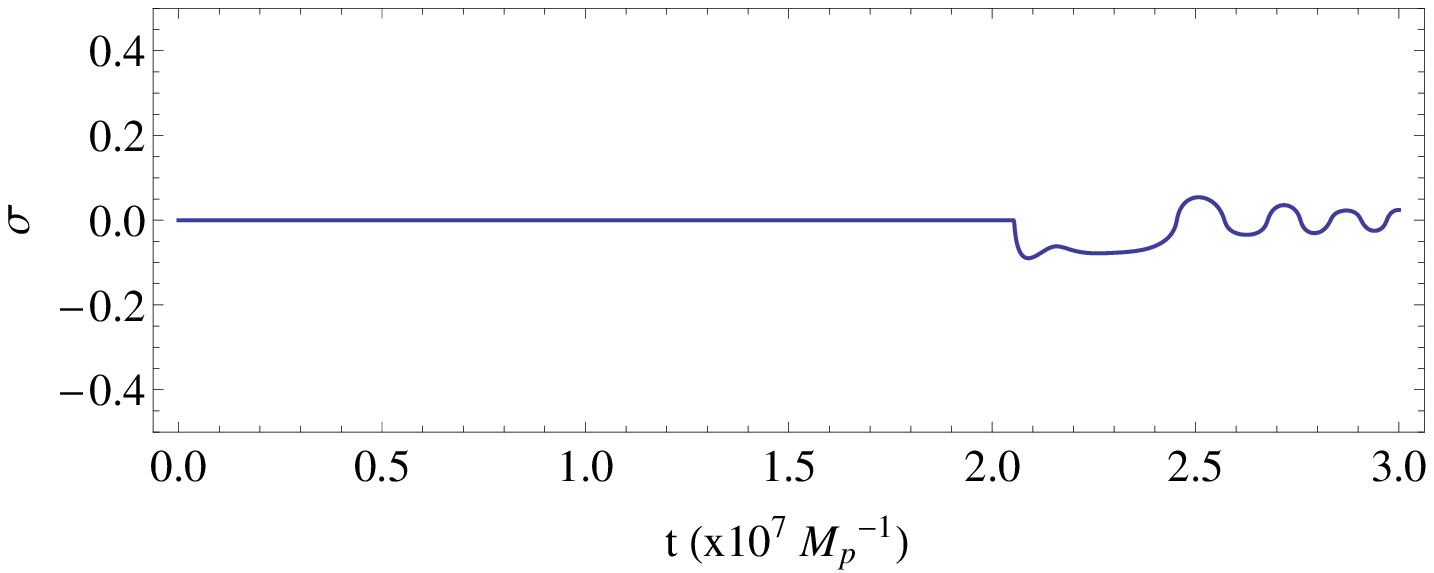}} 
	\scalebox{0.63}{\includegraphics{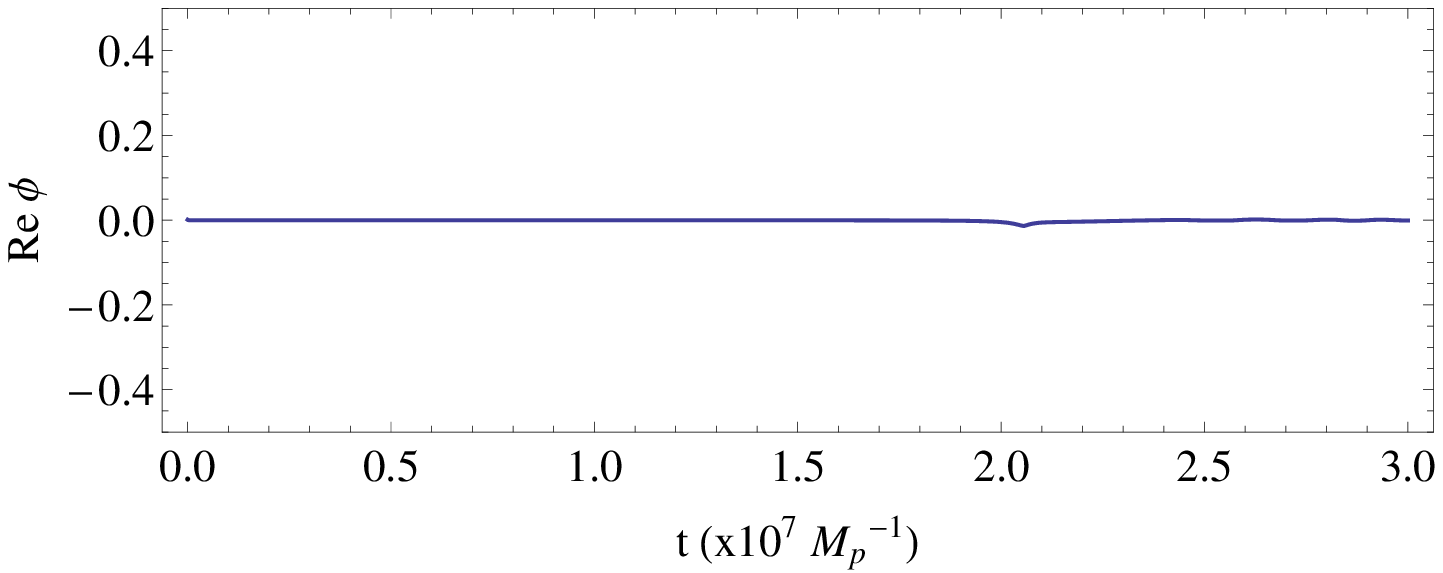}} 
	\scalebox{0.63}{\includegraphics{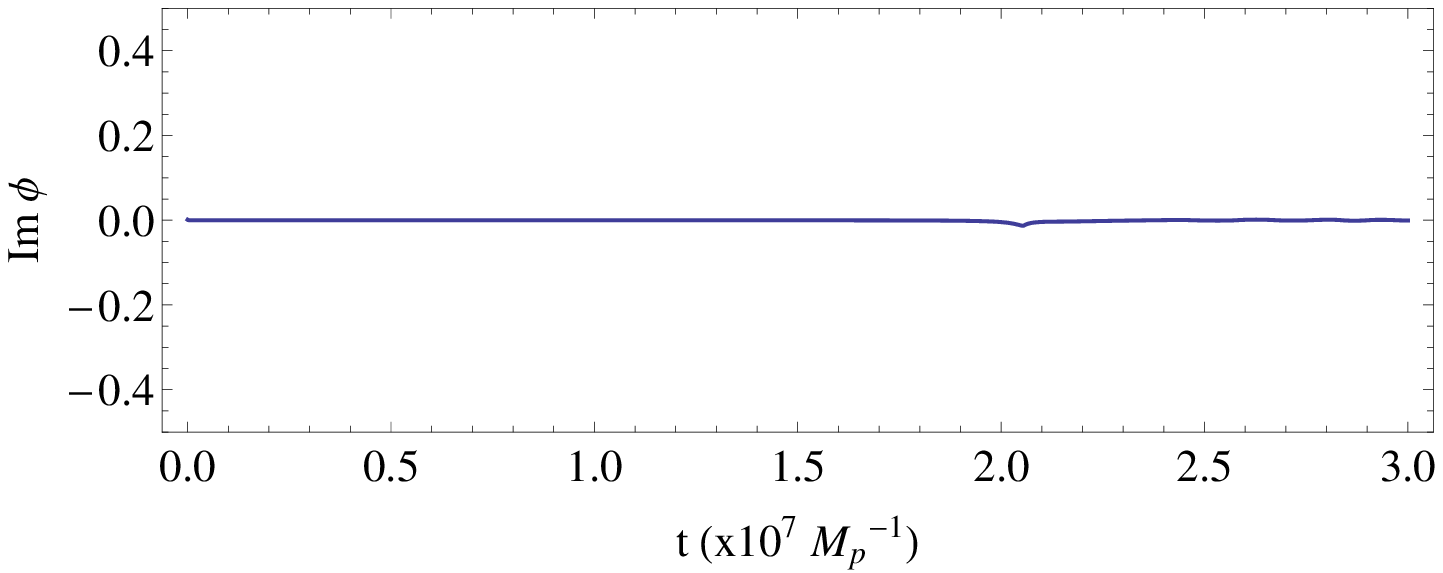}} 
	\scalebox{0.63}{\includegraphics{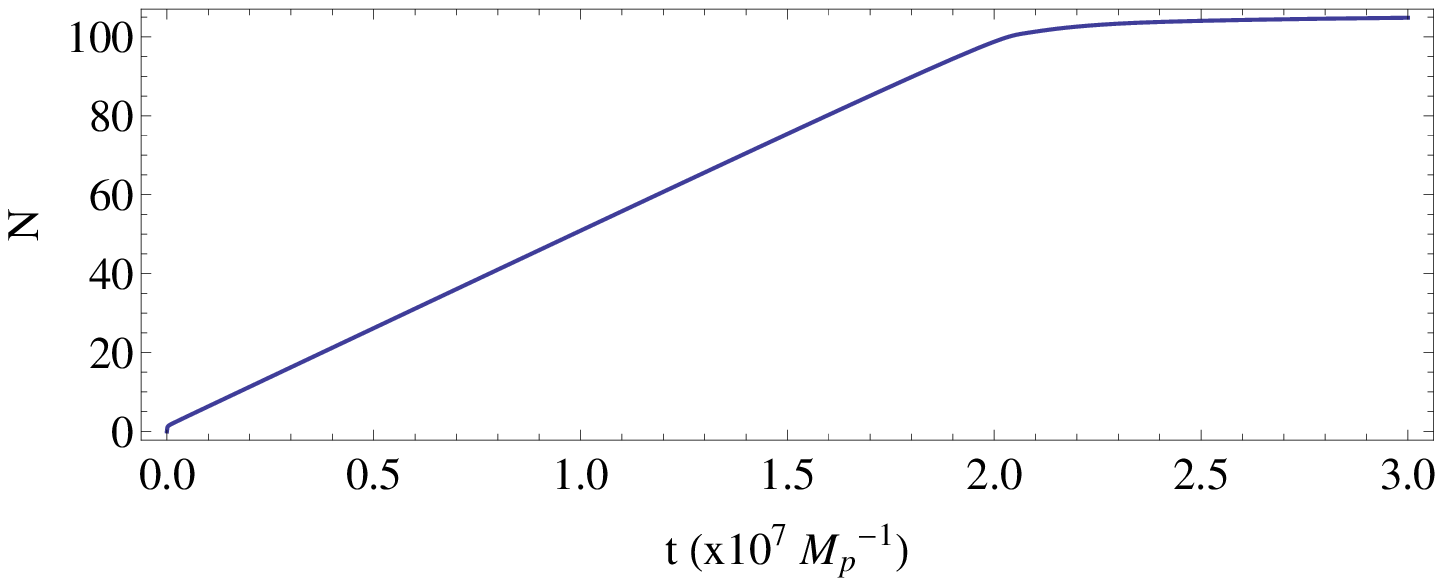}} 
	\caption{\it Numerical solution for $\theta=\pi/2$, $c = 1000$, with initial conditions $\rho_0=6.$, $\sigma_0=0$ and $\phi_0=0.001+0.001i$.}
	\label{fig:sol4}
\end{figure}

As a third example, displayed in Fig.~\ref{fig:sol3}, we show the evolution of the fields and the number of
e-folds for Starobinsky-like initial conditions: $\rho_0=6$, $\sigma_0=0$, $\phi_0=0$, with $\theta=\pi/2$ and $c = 1000$.
The top panel of Fig.~\ref{fig:sol3} shows that, as expected, the inflaton field ($\rho$ in this case)
moves slowly initially, but then accelerates rapidly towards zero as it rolls down the steepening
potential, before exhibiting damped oscillations. The second, third and fourth panels of
Fig.~\ref{fig:sol3} show that, also as expected, the other fields $\sigma, {\rm Re} \, \phi$ and ${\rm Im} \, \phi$
all remain fixed at their minima, and the bottom panel displays the growth in the number of e-folds.
We find for this set of initial conditions that
\begin{eqnarray}
N \; = \; 50: \; \; (n_s, r) & =  & (0.960, 0.004) \, , \nonumber \\
N \; = \; 60: \; \; (n_s, r) & = & (0.967, 0.003) \, .
\label{nsr3}
\end{eqnarray}
Finally, Fig.~\ref{fig:sol4} displays the results of including a small non-zero value of $\phi$ in
the initial conditions: $\rho_0=6$, $\sigma_0=0$, $\phi_0=0.001+0.001i$, for $\theta=\pi/2$ and $c = 1000$. We choose
$|\phi_0|\ll1$ for $\rho>1$ because, as seen in the lower right panel of Fig.~\ref{fig:potpix},
the effective potential rises very steeply as a function of $\phi$ when $\rho$ is large. We see in the
top two panels that $\rho$ and $\sigma$ evolve almost identically as before, whereas the
third and fourth panels show that ${\rm Re} \, \phi$ and ${\rm Im} \, \phi$ exhibit small oscillations
as inflation comes to an end. However, this has negligible effect of the growth in the number
of e-folds, as seen in the bottom panel of Fig.~\ref{fig:sol4}. We find for this case that
\begin{eqnarray}
N \; = \; 50: \; \; (n_s, r) & =  & (0.961,0.004) \, , \nonumber \\
N \; = \; 60: \; \; (n_s, r) & = & (0.968,0.003) \, ,
\label{nsr4}
\end{eqnarray}
results that are very similar to the previous pure Starobinsky case. 

\begin{figure}[!h]
\centering
	\scalebox{0.8}{\includegraphics{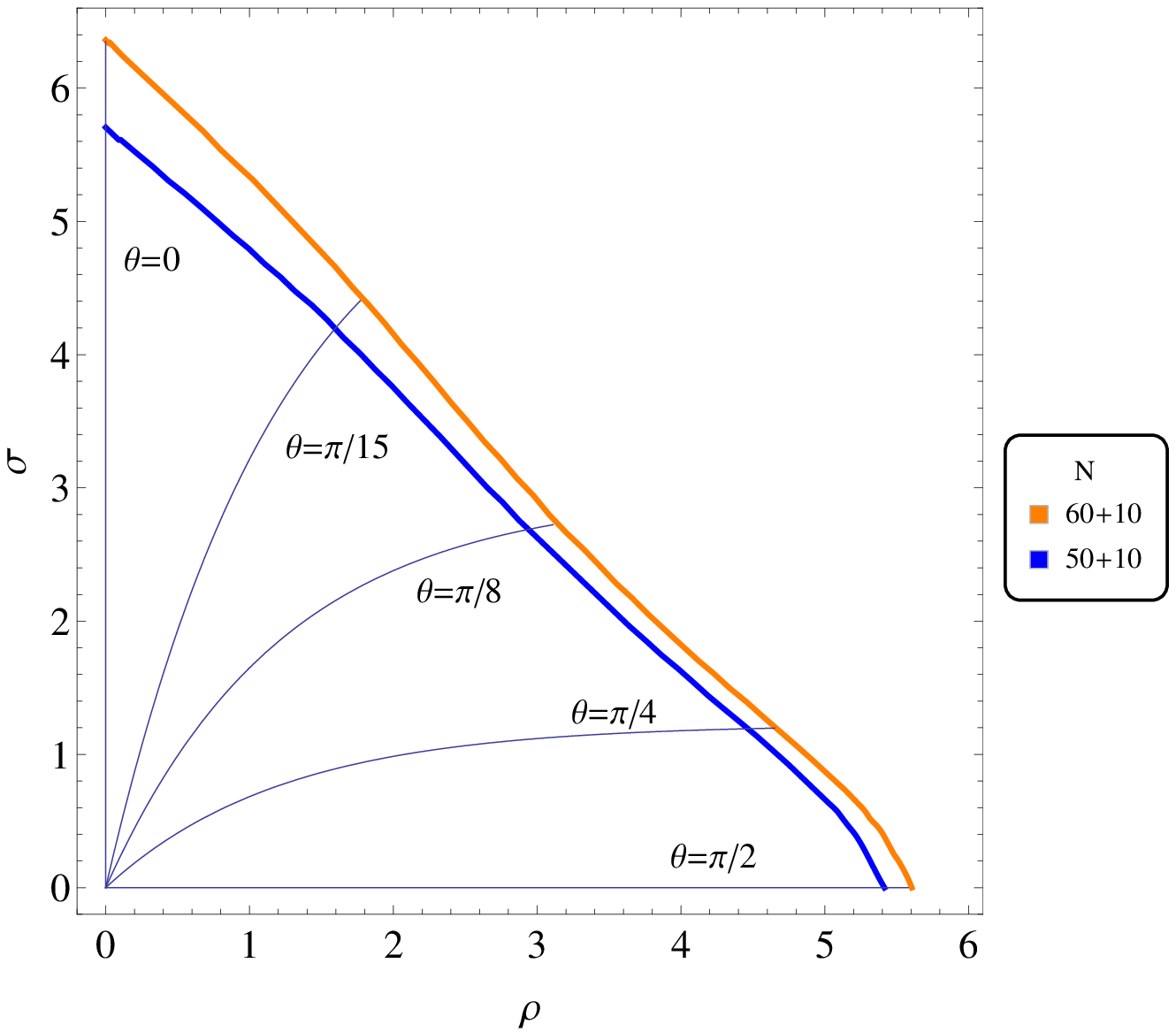}} 
	\caption{\it Parameterization of initial conditions in the $(\rho,\sigma)$ plane for $\phi_0=0$ and $c=200$. 
	} 
	\label{fig:Tplane1}
\end{figure}

\begin{figure}[h!]
\centering
	\scalebox{0.9}{\includegraphics{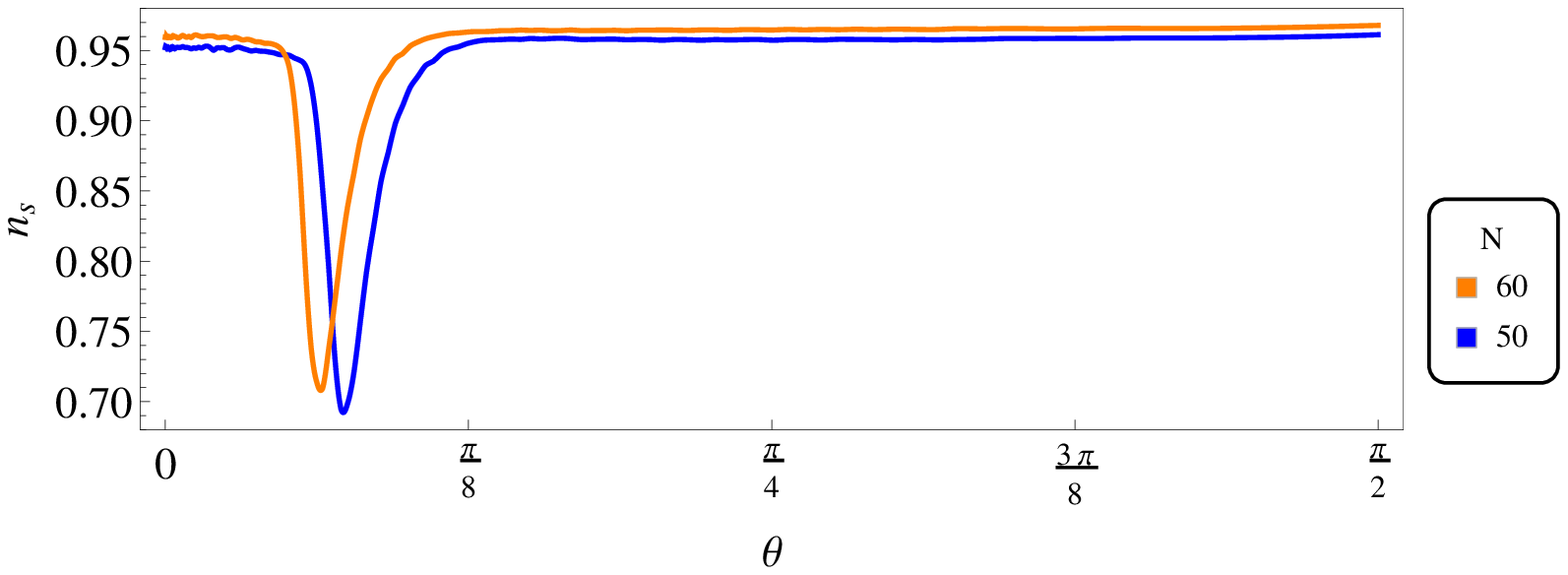}} \\
	\scalebox{0.9}{\includegraphics{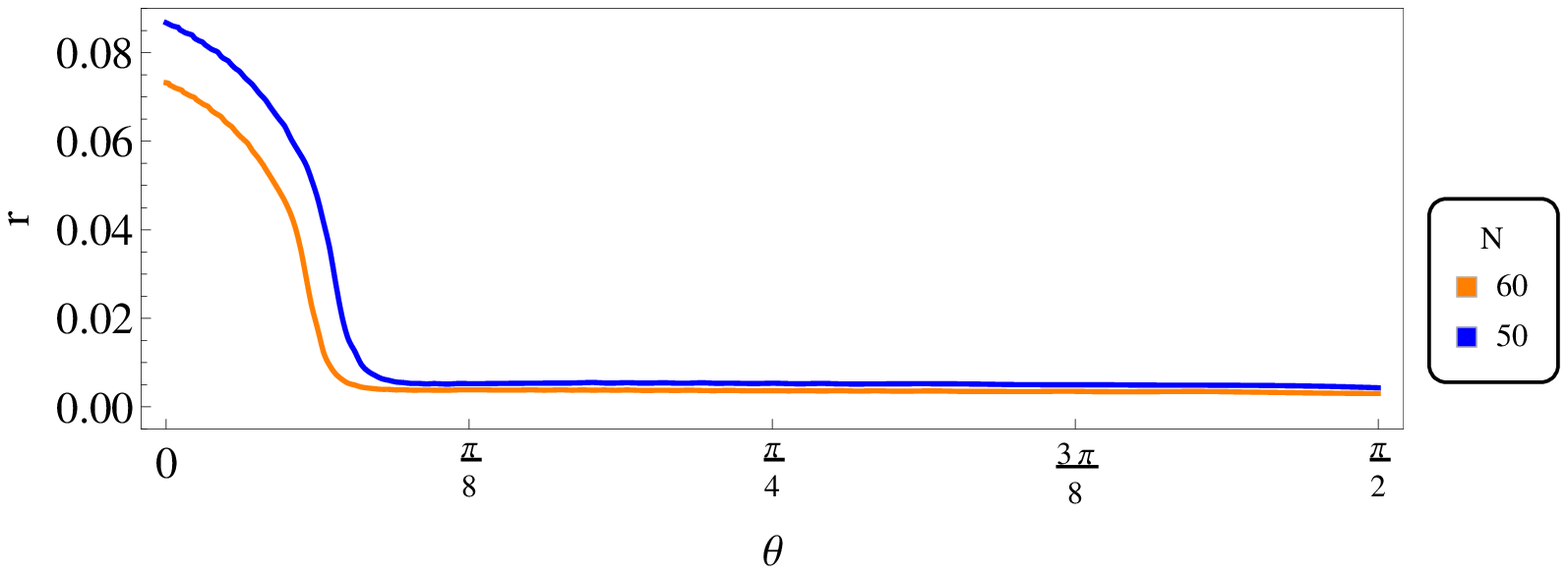}} 
	\caption{\it Upper panel: The scalar tilt as function of $\theta$.
	Lower panel: The tensor-to-scalar ratio as function of $\theta$.}
	\label{fig:nsrth1}
\end{figure}

\begin{figure}[h!]
\centering
	\scalebox{0.9}{\includegraphics{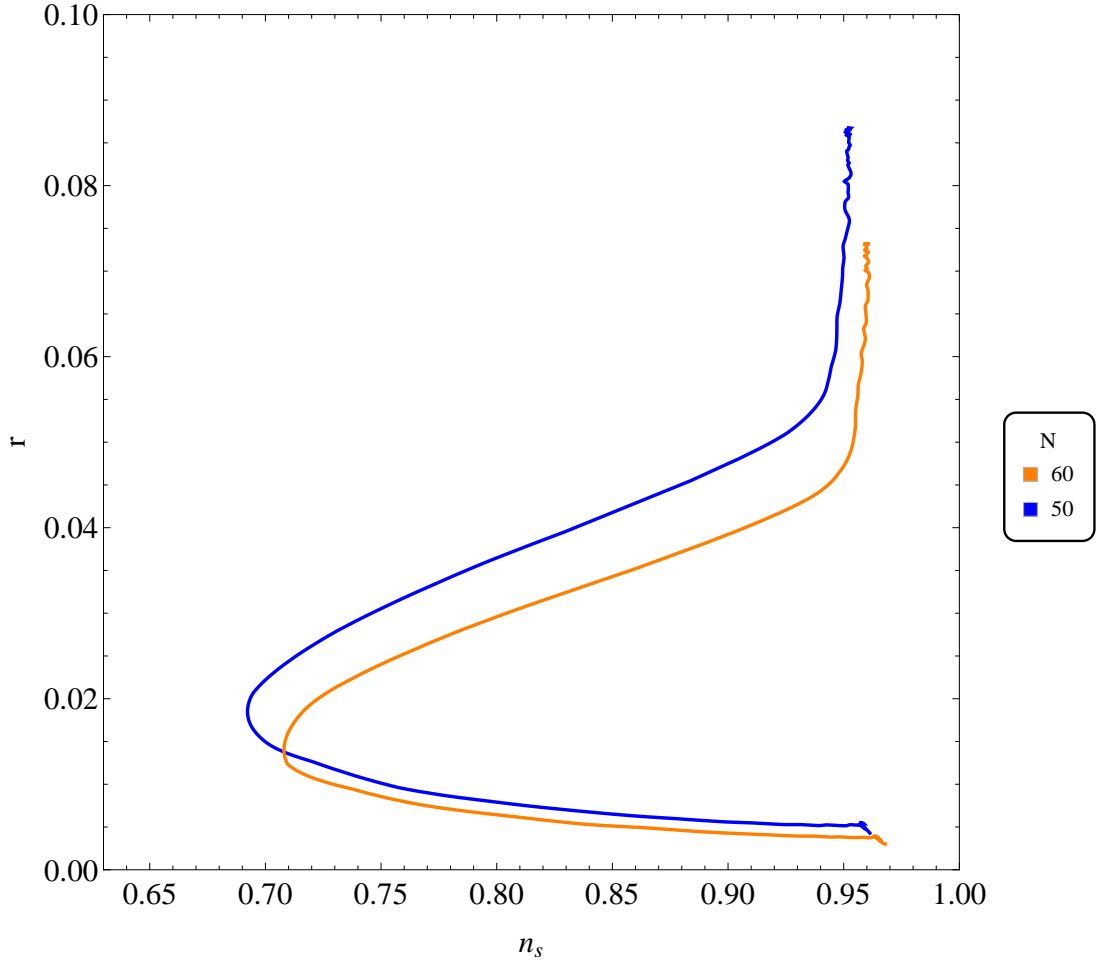}} 
	\caption{\it The parametric curve $(n_s(\theta),r(\theta))$.} 
	\label{fig:nsrth2}
\end{figure}

It is clear from the above results that the scalar tilt and the tensor-to-scalar ratio depend on the initial condition for the complex inflaton field $T$. In order to quantify this dependence more generally, we consider initial conditions in the $(\rho,\sigma)$ plane parametrized by the angle $\theta$, as shown in Fig.~\ref{fig:Tplane1}, restricting our
attention to the case $\phi_0 = 0$. For definiteness we have considered initial conditions on the curve in the $(\rho,\sigma)$ plane that leads to $N+10$ e-foldings of inflation, for $N=50,60$. The resulting $\theta$ dependences of the inflationary
observables $n_s$ and $r$ are displayed in Figs.~\ref{fig:nsrth1} and \ref{fig:nsrth2}.
We see in the upper panel of Fig.~\ref{fig:nsrth1} that $n_s$ is almost independent of
$\theta$, and always within the 68\% CL range favoured by 
WMAP~\cite{WMAP}, Planck~\cite{Planck} and BICEP2~\cite{BICEP2}, except for a region centered around $\theta\sim0.25$. This can be tracked to a sharp enhancement in the power spectrum around these values of $\theta$. In the lower panel of Fig.~\ref{fig:nsrth1} we notice that, as expected,
$r$ decreases monotonically from the large BICEP2-friendly values $r \ga 0.08$
at $\theta = 0$ to the much smaller Planck-friendly values at $\theta = \pi/2$.
We note that the results are symmetric under reflection in the $\rho$ axis:
$\theta \to \pi -  \theta$, and that initial conditions $0 > \theta > - \pi/2$
(and their reflections) would give larger values of $r$ than quadratic inflation
or simply not inflate due to the exponential nature of the potential when $\rho < 0$.
We complete this discussion of the $\theta$ dependence of $r$ and $\theta$
by displaying in Fig.~\ref{fig:nsrth2} the parametric curve $(n_s(\theta),r(\theta))$.

\begin{figure}[h!]
\centering
	\scalebox{0.9}{\includegraphics{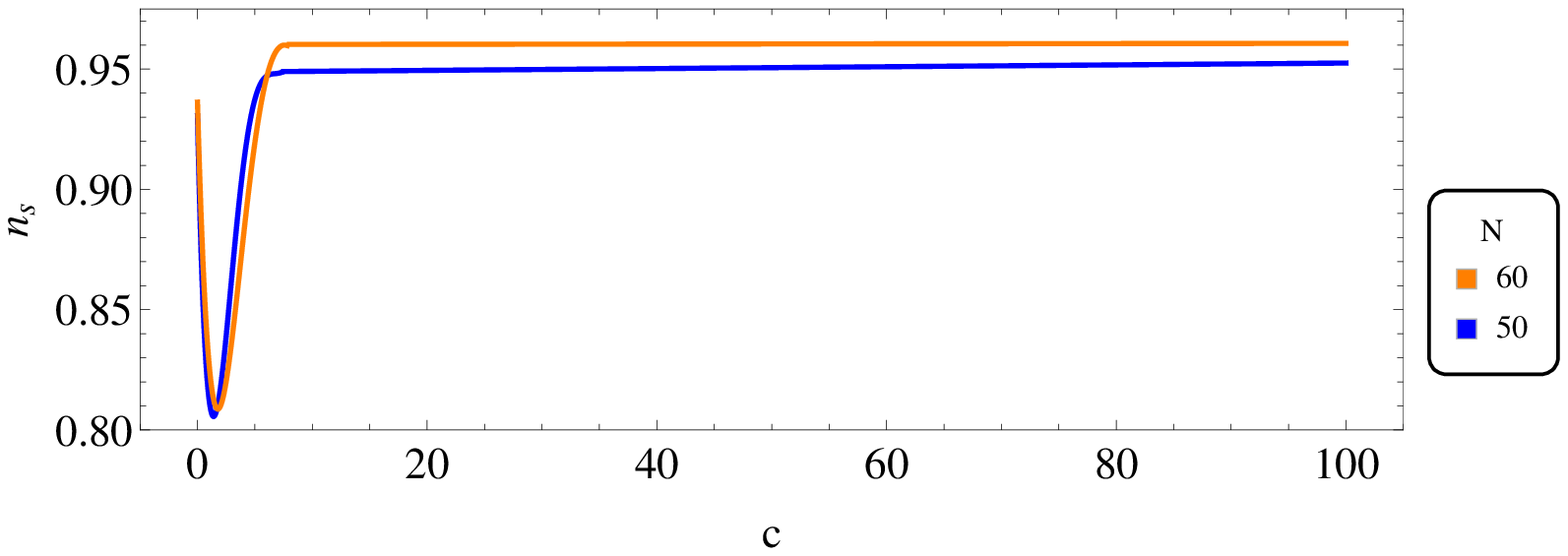}} \\
	\scalebox{0.9}{\includegraphics{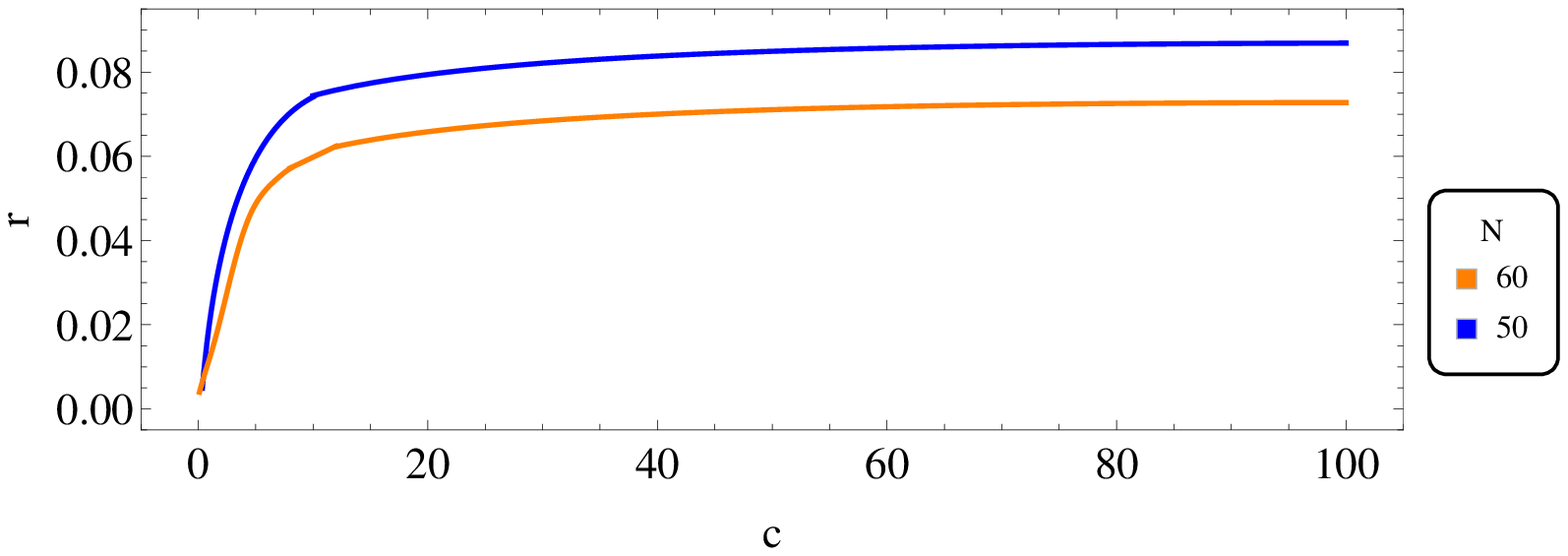}} 
	\caption{\it Upper panel: The scalar tilt as function of the stabilization parameter $c$.
	Lower panel: The tensor-to-scalar ratio as function of $c$. The initial conditions correspond to $\rho_0=0$ and $N_{\rm tot}=N+10$, for $N=50,60$.}
	\label{cdep}
\end{figure}

Up to now, we have fixed the value of $c\gg 1$. Fig.~\ref{cdep}
shows the dependence of $(n_s, r)$ on the constant $c$ for $\rho_0 = 0$. The inital condition $\sigma_0$ is fixed by the requirement of a total of $N+10$ e-foldings; it is dependent on the value of $c$. As one can see, for small $c$, $n_s$ deviates significantly from the range favoured by WMAP, Planck and BICEP2, while for $c\gtrsim10$, it falls within acceptable values. The tensor-to-scalar ratio $r$ rises monotonically and eventually plateaus at the values (\ref{nsr1}) when $c \ga 100$.

\begin{figure}[!h]
\centering
	\scalebox{0.9}{\includegraphics{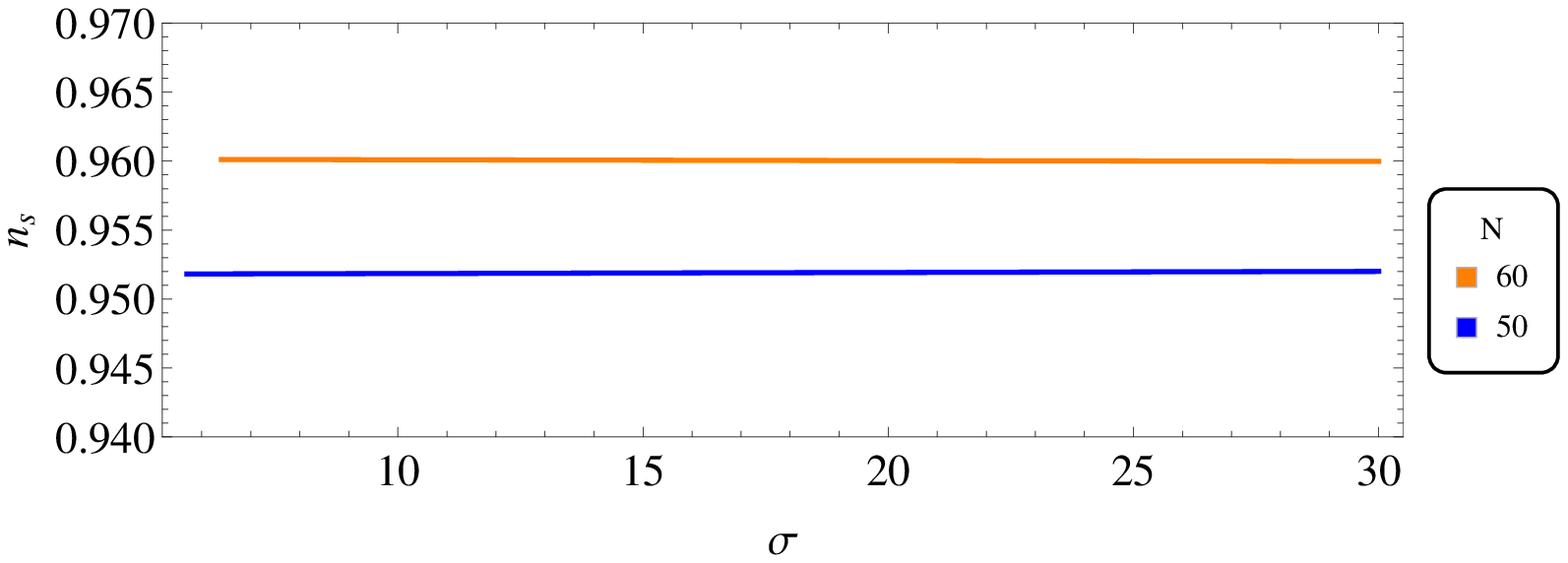}} \\
	\scalebox{0.9}{\includegraphics{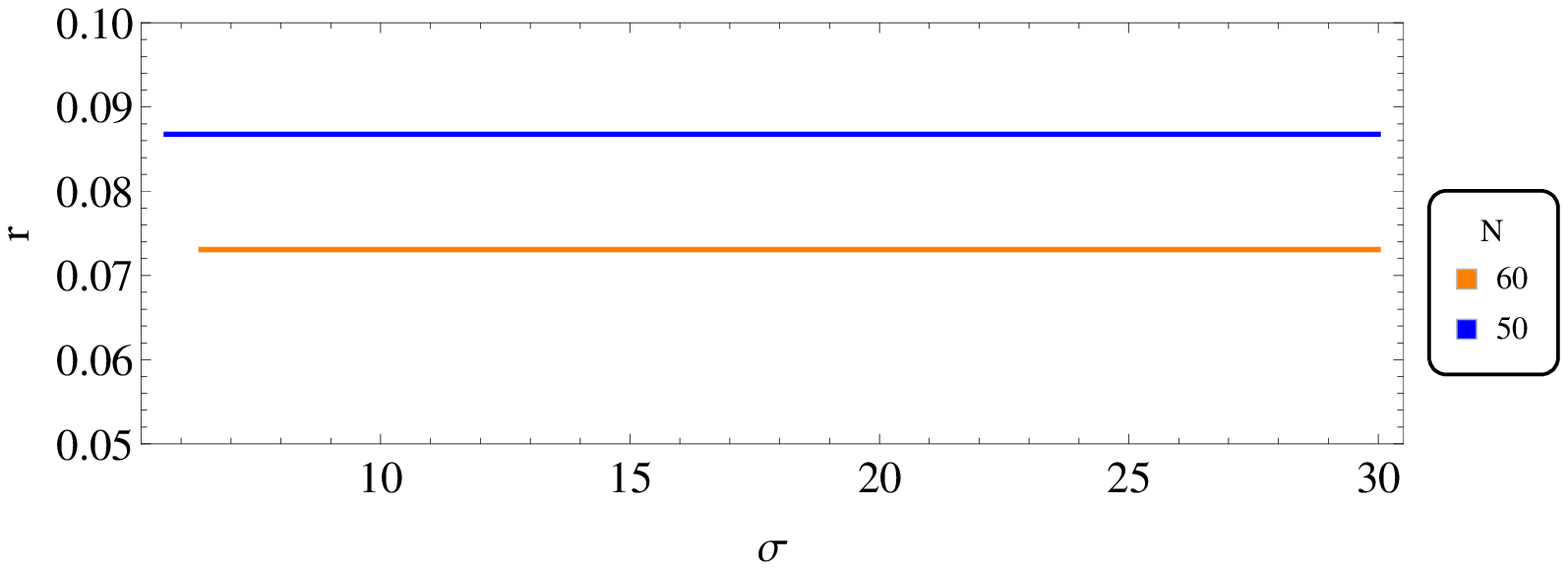}} 
	\caption{\it Upper panel: The scalar tilt $r$ as a function of the initial value of $\sigma$.
	Lower panel: The tensor-to-scalar ratio $n_s$ as a function of the initial value of $\sigma$. Here $\rho_0=0$ and $c=200$.}
	\label{fig:nsrsig}
\end{figure}

So far, we have considered initial conditions for the fields with vanishing time derivatives - a `standing start' - with larger field values than the minimum needed to obtain sufficient e-folds, as reflected in the numbers of e-folds $N = {\cal O}(100)$ in Figs.~2, 4, 5 and 6. In such models, the number of e-folds sufficient to generate the observable universe effectively follow a `rolling start', and the inflationary observables $r$ and $n_s$ may take different values, in general. We display in Fig.~\ref{fig:nsrsig} the dependences of the scalar tilt (upper panel) and the tensor-to-scalar ratio (lower panel) on the initial value of $\sigma$~\footnote{The dependence on the initial value of $\rho$ is equally  insignificant.}. Despite the coupling between the real and imaginary parts of the complex field $T$ which perturbs the field $\rho$, even if it is initially stationary at its minimum $\rho=0$, there is very little dependence of $r$, on the initial value of $\sigma$, as seen in Fig.~\ref{fig:nsrsig}. The scalar tilt is also independent of $\sigma$.

\section{Supersymmetry Breaking}

In the above analysis we have neglected the possible effects of
supersymmetry breaking, which one would expect, in general, to
have little importance for $m_{3/2} \ll m \sim 2 \times 10^{13}$~GeV.
There is certainly plenty of room between this upper limit and the
lower limits imposed by Big Bang nucleosynthesis and the absence of supersymmetric particles
so far at the LHC.

In principle, one could imagine that adding a constant term, $W_0$, to the superpotential (\ref{ourW})
would suffice to induce supersymmetry breaking as in~\cite{ENO8}. However, doing so shifts the minimum
from $\phi = 0, T = 1/2$, very slightly to a supersymmetry preserving AdS vacuum state \cite{klor,lmo,dlmmo} 
as in KKLT~\cite{kklt} and KL~\cite{kl} models. Thus, some form of `uplifting' is necessary and 
the restoration of a Minkowski (or slightly dS) vacuum with supersymmetry breaking is possible
with simple examples of F-term uplifting \cite{lnr,scrucca,dpp,Kallosh:2006dv}.

As a toy example, we consider an unstabilized Polonyi modulus~\cite{pol} as the source of supersymmetry breaking.
Thus, we add to our previous K\"ahler potential (\ref{ourK}) the following terms
\beq
\Delta K \; = \; |Z|^2+\frac{|Y|^2}{(T+\bar{T})^{n}} \, ,
\eeq
where $Z$ is the Polonyi field and the $\{Y\}$ are generic matter fields with unspecified modular weights. We
also add to our superpotential (\ref{ourW}) the terms
\beq
\Delta W \; = \; \mu(Z+\nu) + W(Y) \, .
\label{deltaW}
\eeq
The scalar potential is minimized along the imaginary directions of $T, \phi$ and $Z$ for
\beq
{\rm Im}\,\phi \; = \; {\rm Im}\, Z \; = \; \sigma \; = \; 0 \, .
\eeq
Along the real directions, the minimization must be performed numerically in general, 
since it depends on the value of $\mu$. However, for $\mu\ll m$, the conditions
\beq
\partial_{{\rm Re}\,\phi}V \; = \; \partial_{{\rm Re}\,Z}V \; = \; \partial_{\rho}V \; = \; V \; = \; 0
\label{conditions}
\eeq
for the minimum yield, to second order in $\mu/m$, in Planck units,
\beq
{\rm Re}\,\phi\simeq \sqrt{3}\,\frac{\mu}{m}\ , \ \ \rho\simeq \sqrt{6}(1-\sqrt{3})\left(\frac{\mu}{m}\right)^2\ , \ \ {\rm Re}\,Z\simeq -1+\sqrt{3}\ , \ \ \nu\simeq 2-\sqrt{3} \, .
\eeq
We explicitly see that the shifts in ${\rm Re}\,\phi$ and $\rho$ induced by the parameter $\mu$
are small when $\mu \ll m$. Finally, we recall that since
\beq
D_{Z}W\simeq \sqrt{3}\mu \, ,
\eeq
supersymmetry is broken with
\beq
m_{3/2}\simeq \mu \, ,
\eeq
and the induced masses, trilinear and bilinear terms for the matter fields $\{Y\}$ correspond to
(after a constant rescaling of the superpotential $W \to e^{\sqrt{3}-2}W$
\beq
m_0 \simeq m_{3/2} \ , \ \ A_0\simeq (3-\sqrt{3}) m_{3/2}\ , \ \ B_0\simeq (2-\sqrt{3}) m_{3/2} \, ,
\eeq
as in models of minimal supergravity \cite{bfs}.
The dependence of these parameters on the modular weight $n$ appears at higher order in $(\mu/m)$.

In order to avoid the well-known problems associated with the minimal Polonyi model~\cite{prob},
we can extend this analysis by considering stabilization~\cite{dlmmo,Kallosh:2006dv,dine} of the Polonyi field via the K\"ahler potential 
\beq
\Delta K \; = \;   |Z|^2-\frac{|Z|^4}{\Lambda^2}+\frac{|Y|^2}{(T+\bar{T})^{n}}
\eeq
with the same superpotential (\ref{deltaW}).
In this case, the scalar potential is also minimized along the imaginary directions of $T,\phi$ and $Z$ with
${\rm Im}\,\phi = {\rm Im}\, Z = \sigma =0$.

Once again, along the real directions, the minimization must be performed numerically
since it is dependent on the values of both $\mu$ and $\Lambda$. 
For $\Lambda\ll1$ and $\mu\ll m$, where $m$ is the inflaton mass, the conditions
(\ref{conditions}) can now be solved approximately to give
\beq
{\rm Re}\,\phi\simeq\frac{\mu}{m}\ , \ \ \rho\simeq -2\sqrt{\frac{2}{3}}\left(\frac{\mu}{m}\right)^2\ , \ \ {\rm Re}\,Z\simeq \frac{\Lambda^2}{\sqrt{12}}\ , \ \ \nu\simeq \frac{1}{\sqrt{3}}
\eeq
with higher order terms at most $\mathcal{O}(\frac{\mu}{m}\Lambda^2)$.
Since
\beq
D_{Z}W\simeq \mu \, ,
\eeq
supersymmetry is broken, with
\beq
m_{3/2}\simeq \frac{\mu}{\sqrt{3}} \, ,
\eeq
and the induced masses, bilinear and trilinear terms for the matter fields $\{Y\}$ are
\beq
m_0 \simeq m_{3/2} \ , \ \ A_0\simeq 0\ , \ \ B_0\simeq-m_{3/2} \, ,
\eeq
as in minimal supergravity models with vanishing $A$ terms or models of pure gravity mediation \cite{eioy}.
In this case the dependence on the modular weight $n$ appears at $\mathcal{O}(\frac{\mu}{m}\Lambda^2)$.

\section{Summary and Conclusions}

We have proposed in this paper a simple two-field no-scale supergravity model of inflation whose
predictions for the scalar-to-tensor perturbation ratio $r$ interpolate between limits
that are BICEP2-friendly and Planck-friendly, Starobinsky-like: $0.09 \ga r \ga 0.003$. As we have shown, this model also
yields $n_s \sim 0.96$ in most of field space, as indicated by WMAP, Planck and BICEP2 data. 
Our model is
based on the form of effective low-energy field theory derived from orbifold compactifications
of string theory, and can accommodate a Polonyi mechanism for supersymmetry breaking
that is suitable for particle phenomenology.


We await with interest confirmation of the B-mode polarization measurement made by BICEP2, and verification that is mainly of primordial origin. In the mean time, we note that over a region of its parameter space our no-scale model yields values of $r$ which may be compatible at the 68\% CL with each of the WMAP, Planck and BICEP2 measurements. In that sense, our model may indeed `fit them all'.

\section*{Acknowledgements}

We thank Krzysztof Turrzy\'nski for emphasizing the importance of two-field effects in the absence of
a stabilizing term in the K\"ahler potential.
The work of J.E. was supported in part by the London Centre for Terauniverse Studies
(LCTS), using funding from the European Research Council via the Advanced Investigator
Grant 267352 and from the UK STFC via the research grant ST/J002798/1. J.E. also thanks
Hong-Jian He for his kind hospitality at the Center for High Energy Physics,
Tsinghua University, and S.-H. Henry Tye for his kind hospitality at the Institute for Advanced Study, 
Hong Kong University of Science and Technology, while this work was being carried out.
The work of D.V.N. was supported in part by the DOE grant DE-FG03-
95-ER-40917 and would like to thank A.P. for inspiration. The work of M.A.G.G. and
K.A.O. was supported in part by DOE grant DE-FG02-94-ER-40823 at the University of
Minnesota.

\end{document}